\documentstyle[aps,epsfig,preprint,multicol]{revtex}
\newcommand{\sgn}{{\,\rm sgn\,}}
\newcommand{\str}{{\,\rm str\,}}
\newcommand{\strln}{{\,\rm str\,\ln}}
\newcommand{\GL}{{\rm GL}(1|1)}
\newcommand{\D}{{\bf D}}
\newcommand{\A}{{\bf A}}
\begin{document}

\draft
\title{Spectral and Transport Properties of 
  Quantum Wires with Bond Disorder}
\author{Alexander Altland${}^a$ and  Rainer Merkt${}^{b}$}
\address{${}^a$Theoretische Physik III, Ruhr-Universit\"at-Bochum,
  44780 Bochum, Germany\\ ${}^b$ Institut f\"ur theoretische Physik,
  Z\"ulpicher Str. 77, 50937 K\"oln, Germany}
\maketitle                                 
\begin{center}
  \it Dedicated to E. M\"uller-Hartmann on the occasion of his 60th birthday.
\end{center}
\begin{abstract}
  Systems with bond disorder are defined through lattice Hamiltonians
  that are of pure nearest neighbour hopping type, i.e.  do not
  contain on-site contributions. They stand representative for the
  general family of disordered systems with chiral symmetries.
  Application of the Dorokhov-Mello-Pereyra-Kumar transfer matrix
  technique [P. W. Brouwer {\it et al.}, Phys. Rev.  Lett {\bf 81},
  862 (1998); Phys. Rev. Lett. {\bf 84}, 1913 (2000)] has shown that
  both spectral and transport properties of quasi one-dimensional
  systems belonging to this category are highly unusual. Most notably,
  regimes with absence of exponential Anderson localization are
  observed, the single particle density of states exhibits singular
  structure in the vicinity of the band centre, and the manifestation of
  these phenomena depends in an apparently topological manner on the
  even- or oddness of the channel number. In this paper we re-consider
  the problem from the complementary perspective of the non-linear
  $\sigma$-model.  Relying on the standard analogy between
  one-dimensional statistical field theories and zero-dimensional
  quantum mechanics, we will relate the problem to the behaviour of a
  quantum point particle subject to an Aharonov-Bohm flux.  We will
  build on this analogy to re-derive earlier DMPK results, identify a
  new class of even/odd staggering phenomena (now dependent on the
  total number of sites in the system) and trace back the anomalous
  behaviour of the bond disordered system to a simple physical
  mechanism, viz. the flux periodicity of the quantum Aharonov-Bohm
  system. We will also touch upon connections to the low energy
  physics of other lattice systems, notably disordered chiral systems
  in $0$ and $2$ dimensions and antiferromagnetic spin chains.
\end{abstract}\vspace{0.5cm}

\section{Introduction}
\label{sec:intro}

In 1979, Wegner and Oppermann\cite{oppermann79} introduced a
variant of the disordered lattice Anderson model they termed
'sublattice system'.  The sublattice system differs from the generic
Anderson model in that its Hamiltonian does not contain on-site matrix
elements, i.e.  is of {\it pure} hopping type.
For a long time this species of disordered electronic systems went
largely unnoticed. The status rapidly
changed\cite{gade93,miller96,BMSA:98,altland99:NPB_flux,fukui99,furusaki99,biswas,guruswamy00,brouwer00:_nonun,brouwer00_off,brouwer:trans_dos_off,favrizio}
after two aspects became generally appreciated: first, models with
sublattice structure occur in a number of physical applications. The
random flux model, lattice QCD models\cite{verbaarschot00}, random
antiferromagnets\cite{fisher94:_random}, models of gapless
semiconductors\cite{ovchinikov77} and effective models of transport in
manganese oxides\cite{mueller-hartmann96} are of sublattice type or at least
acquire sublattice structure in limiting cases.  Second, and contrary
to naive expectations based on universality and insensitivity to
details of the microscopic implementation of disorder, the low energy
properties of the sublattice system differ drastically from those of
the generic Anderson model:
\begin{itemize}
\item In contrast to the Anderson model, average spectral and
  transport properties of the sublattice system sensitively depend on
  the value of the Fermi energy, $E_F$. For Fermi energies far away
  from the centre of the tight binding band, $E_F = 0$, the sublattice
  system falls into the universality class of standard disordered
  electron systems (which simply follows from the fact that a Fermi
  energy $E_F \not=0$ can be interpreted as a constant non-vanishing
  on-site contribution to the Hamiltonian.) However, in the vicinity
  of zero energy drastic deviations from standard behaviour occur:
\item In dimensions $d \le 2$, the average density of states (DoS),
  $\nu(E)$, exhibits singular behaviour upon $E$ approaching the band
  centre.
\item Perturbative one-loop RG calculations\cite{gade93} as well as
  the analysis of Ref.\cite{guruswamy00} indicate that right in the
  middle of the band the $2d$ system is metallic, i.e. does not drive
  towards an exponentially localized phase.
\item Several phenomena of apparent topological origin are observed.
  E.g., the DoS profile of sublattice quantum dots (systems in the
  'zero-dimensional' limit) sensitively depends on the total number of
  sites of the host lattice being even or odd. Similarly, the
  properties of quasi one-dimensional sublattice systems depend on the
  number of channels, $N$, being even or odd. For $N$ even the
  transport behaviour is conventional -- conductance exponentially
  decreasing on the scale of a certain localization length $\xi$ --
  whereas the energy dependent DoS vanishes in a close to linear
  fashion. In contrast, for $N$ odd, the wire at $E=0$ is metallic,
  i.e. the length-dependent conductance decays only algebraically. At
  the same time, the DoS diverges upon approaching the band centre. 
\end{itemize}
A schematic summary of the band centre phenomenology of sublattice
models is displayed in table~\ref{tab:1}. 

All these phenomena root in the fact that the sublattice Hamiltonian
possesses a discrete symmetry not present in the Anderson model:
$[\sigma_3,\hat H]_+=0$, where $[\;,\;]_+$ is the anti-commutator and
$\sigma_3$ a site-diagonal operator that takes values $+1/-1$ on
alternating sites. The presence of this 'chiral' symmetry implies that
sublattice systems fall into symmetry classes different from the three
standard Wigner-Dyson classes 'unitary', 'orthogonal' or 'symplectic'.
To be specific, let us focus on the simplest case of a sublattice
system with broken time reversal invariance but good spin rotation
symmetry (the analogue of the Anderson model of unitary symmetry.)
Chiral systems fulfilling these two extra symmetry criteria belong to
a symmetry class denoted $A$III in the terminology of
Ref.\cite{zirnbauer96}. An alternative denotation, coined in the
paper\cite{verbaarschot94:RMT2}, is ChGUE for 'Chiral GUE'.
\begin{figure}[hbt]
  \begin{center}
    \begin{tabular}{|c||c|c|c|c|c|}\hline
&\multicolumn{2}{c|}{$0d$}&\multicolumn{2}{c|}{$1d$}&$2d$\cr\hline
staggering &$L$ even&$L$ odd& $N$ even &$N$ odd&\cr\hline\hline
Conductance ($\epsilon=0$)
&vanishing& non-vanishing&  localization &  deloc.& deloc. 
\cr\hline
Density of states
&spectral gap & zero energy states
&spectral gap & divergence & divergence\cr\hline
    \end{tabular}\\[0.5cm]
    \caption{Schematic overview over phenomenology of sublattice
      systems. $N:$ number of channels of a quasi one-dimensional
      quantum wire, $L:$ number of sites of a sublattice quantum dot.}
    \label{tab:1}
  \end{center}
\end{figure}

As with conventional disordered electronic systems of Wigner-Dyson
symmetry, universal transport and thermodynamic properties of systems
of class $A$III can be described in terms of effective low energy
theories. E.g., the results for sublattice quantum wires summarized
above have been obtained within a symmetry adapted variant of the
Dorokhov-Mello-Kumar-Pereyra (DMPK) transfer matrix
approach\cite{BMSA:98,brouwer00_off,brouwer00:_nonun}. This theory
differs from the standard cases of unitary symmetry in that the
transfer matrices describing the propagation through the system take
values in a different target space.  Similarly, the general field
theory approach to disordered electronic systems, the nonlinear
$\sigma$-model, has been extended to systems of class
$A$III\cite{gade93,altland99:NPB_flux,favrizio,fukui99}, too.  Like
its conventional relatives, the $A$III variant of the $\sigma$-model
is a matrix field theory whose internal structure depends on the
specific implementation (boson replicas, fermion replicas or
supersymmetry.)  Previous studies of these models have focused on the
two-dimensional case\cite{gade93,guruswamy00}, where information on
long range behaviour can be obtained from renormalization group
calculations, or on the zero-dimensional
case\cite{verbaarschot96:susy} where the model can be evaluated
rigorously by full integration over the zero-mode manifold.

It is the purpose of the present paper to analyse the intermediate,
quasi one-dimensional variant of the field theory.  As mentioned above
key aspects of the phenomenology of quasi one-dimensional sublattice
quantum wires have been discussed previously within the framework of
the DMPK approach, and it is near at hand to ask what motivates
revisiting the problem. Referring for a more substantial discussion to
section \ref{sec:summary-results} below, we here merely mention a lose
collection of points. The field theory approach enables one to
approach the problem from a comparatively broad perspective.
Specifically, the one-dimensional variant of the model is but a
representative of a larger family of 'chiral' $\sigma$-models. This
makes possible to relate the behaviour of the one-dimensional system
to the extensively studied zero and two dimensional cases. Further,
the Green function oriented $\sigma$-model formalism enables one to
'microscopically' implement coupling operators connecting the wire to
external leads.  (Within previous DMPK formulations, the coupling has
been treated somewhat implicitly, see however\cite{brouwer}.)
Unexpectedly, we will observe strong, albeit universal sensitivity to
the modelling of the coupling and yet another class of staggering
phenomena.  The origin of these effects, and their connection to the
channel number staggering mentioned before will be discussed below.
Finally, the $\sigma$-model of the time reversal non-invariant
sublattice is the by far most simple of all ten\cite{zirnbauer96}
nonlinear $\sigma$-models of disordered systems: it has only four
degrees of freedom, two Grassmann and two ordinary integration
variables, the minimal set needed to construct a supersymmetric matrix
model. This makes it an ideal tutorial system on which generic
properties of the field theory approach to disordered quantum wires
can be studied. We have tried to pedagogically expose several of these
aspects, particularly the analogy one-dimensional field theories vs.
point-particle quantum mechanics which plays an all important role in
the present context.  Nonetheless, the analysis below will be at times
technical. To make its results and various qualitative connections
generally accessible, the following section contains a synopsis of the
paper.

\subsection{Summary of Results and Qualitative Discussion}
\label{sec:summary-results}

\begin{figure}[htbp]
  \begin{center}
    \epsfig{file=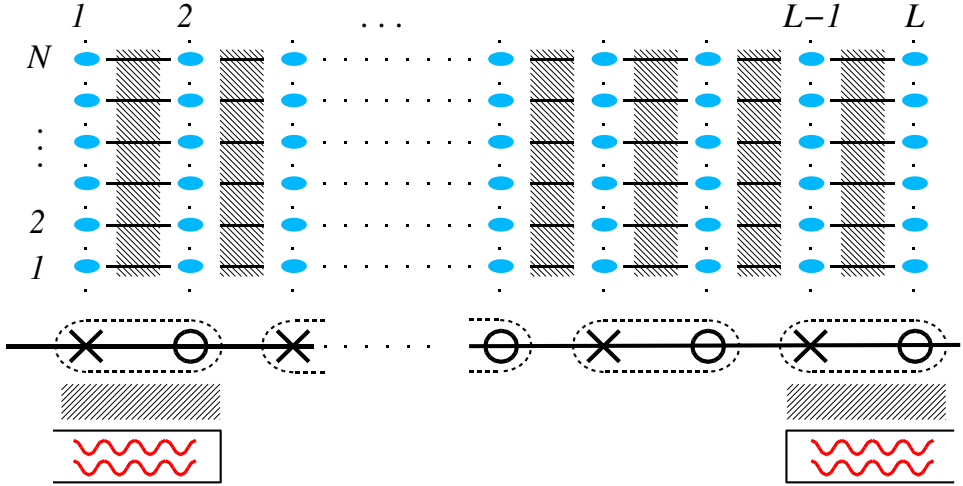,height=6.5cm} \\[0.5cm]
    \caption{Quasi one dimensional sublattice system. The system is
      realized in terms of a sequence of sites, alternatingly
      designated by crosses and circles, each of which carries $N$
      states (the ovals). Hopping between the sites is mediated by a
      regular Hamiltonian (the lines connecting the ovals) plus a weak
      random contribution (hatched areas). At either end, a number of
      sites is coupled to continuum states (the wavy lines)
      propagating in leads.}
    \label{fig:figure1}
  \end{center}
\end{figure}

Consider the system depicted in figure Fig.~\ref{fig:figure1}: a
sequence of $L$ sites (alternatingly designated by crosses and
circles) each of which supports $N$ electronic states, or orbitals
(represented by the vertical stacks of ovals.) Nearest neighbour
hopping is controlled by a regular tight binding contribution,
diagonal in the orbital index, (the horizontal line segments) plus
some bond randomness that connects different orbitals (the hatched
areas). As in Refs. \cite{BMSA:98}, we allow for some 'staggering' in
the tight binding amplitudes, i.e. the hopping amplitudes regularly
alternate in strength (the alternating bond length.) At either end, a
number of sites is coupled through some tunnelling barriers (horizontal
hatched areas) to leads.

In this paper, three different  regimes will be considered: (i) the
'quantum dot' case, defined through the criterion that the time to
diffusively propagate through the system is shorter than the Golden
rule escape time into the leads, (ii) a diffusive regime, where the
order of these time scales is inverted but localization effects do not
yet play a role, and (iii) the regime of long systems with pronounced
Anderson localization. 

Beginning with the quantum dot case (i), we find that both transport and
spectral properties sensitively depend on the number of sites $L$
being even or odd. Specifically, the DoS exhibits a singular spike in
the band centre if the number of sites is odd (and the coupling to the
leads switched off). Away from zero energy, $\nu(\epsilon)$ is
strongly suppressed up to values $\nu \sim N\Delta$, where $\Delta$ is
the mean level spacing. The conductance equals $\gamma N/2$, where
$\gamma$ is a measure for the strength of the lead coupling, as for
conventional quantum dots. In contrast, for an even number of sites,
the DoS shows the spectral 'microgap' of width $\Delta$ characteristic
for finite random systems with chiral symmetries\cite{verbaarschot00}.
Curiously, the conductance {\it vanishes} provided that only one site
at either end is connected to the leads. The strong sensitivity to the
total number of sites in the system disappears if more than one site
in the terminal regions is coupled to leads (cf. Fig.
\ref{fig:figure1}.) Summarizing, the tendency of sublattice systems to
exhibit staggering phenomena pertains to the zero-dimensional limit.
However, unlike in the localized regime, the relevant integer control
parameter is the number of sites $L$ and not the number of channels
$N$. A qualitative interpretation of these phenomena will be given
below.

In this paper, only limited attention will be payed to the
intermediate diffusive regime (ii). It is likely that the staggering
phenomena observed in the zero-dimensional regime will have
interesting, albeit non-universal extensions into the diffusive
regime.  We here avoid the confrontation with these effects by
connecting several sites to the leads. This leads to equilibrated
behaviour with Ohmic conductance, as for ordinary wires.  As for the
density of states, we have explored the influence of spatially
fluctuating diffusion modes on the spectral microgaps discussed above.
(Semiclassically, the spectral gaps observed in chiral or
superconductor systems can be interpreted as due to an accumulation of
diffusion modes (see Ref.  \cite{altland99:NPB_flux} for a detailed
discussion of this point).  For large energies $\epsilon \gg \Delta$,
these modes can be treated perturbatively by standard diagrammatic
methods.) Surprisingly, no corrections are found up to and including
three loop order. This is a speciality of the one-dimensional case.
For a two dimensional system diffusive modes would lead to a
modulation of the spectrum on the scale of the Thouless energy.

In the localized regime we reproduce the results found earlier within
the DMPK approach: for an even number of channels, the conductance
decays exponentially on the scale of a localization length $\xi \propto
N l$, where $l$ is the mean free path. In contrast, for an odd number
of channels algebraic scaling, $g\sim L^{-1/2}$, supported by one
delocalized mode is observed\cite{BMSA:98}. A comparably strong
even/odd dependence is observed in the behaviour of the DoS. For $N$
even, the DoS vanishes at zero energy as $\nu(\epsilon) \sim -|s|
\ln(|s|)$, where $s=\pi \epsilon/\Delta_\xi$, and the characteristic
scale of the gap, $\Delta_\xi$, is the level spacing of a system of
length $L\sim \xi$.  This behaviour has a simple interpretation:
roughly speaking, a system with $L\gg \xi$ can be viewed as a sequence
of $L/\xi$ decoupled 'localization volumes'.  On small energy scales
$\epsilon < \Delta_\xi$, dynamics within each of these is
approximately ergodic.  One would then expect the DoS to be gapped, as
for the sublattice quantum dot, with $\Delta_\xi$ as the
characteristic level spacing.  Generalizing an earlier
result\cite{theodorou76,eggarter78} for the specific case $N=1$, it has
been found that for an odd number of channels the DoS diverges as
$\nu(\epsilon) \sim 1/ (|s|\ln^3|s|)$\cite{brouwer00_off}. This
accumulation of spectral weight can be interpreted in two different
directions. First, it is near at hand to view the algebraically
decaying conductance observed in the odd case as a resonant tunnelling
phenomenon: the principal tendency to localize is outweighed by the
high density of states in the vicinity of zero energy. (This picture
was first suggested by V.  Kravtsov.)  Second, it is tempting to
interpret the zero energy peak as an generalization of the singular
spike found in the $L$-odd zero-dimensional case.  Unfortunately, we
are not aware of a qualitative picture explaining this analogy. At
least technically, however, both phenomena can be traced back to a
common origin. Finally, a periodic modulation, or staggering, of the
hopping amplitudes can be employed to continuously interpolate between
the $N$-even and $N$-odd case, respectively.

We next briefly outline the field theory route to exploring the above
phenomenology. Within the fieldtheoretical approach, long range
properties of the system are described by a functional integral with
action
\begin{equation}
  \label{eq:46}
S = \int_0^L dr \left[-{\xi\over 8} \str(\partial T\partial T^{-1})
  - {N+f\over 2} \str(T\partial T^{-1}) -i {s\over 2} \str(T+T^{-1})\right] +
S_{\rm Gade} + S_{\rm T}.  
\end{equation}
Here we have introduced a continuum variable $r \in [0,L]$
replacing the formerly discrete site index, $f$ is a parameter related
to the staggering strength, $s=\epsilon \pi \nu$, where $\nu$ is the
bulk metallic density of states. Further, $T$ is a matrix field taking
values in $\GL$, i.e. the group of two-dimensional invertible
supermatrices, and 'str' is the supertrace. Finally, $S_{\rm
  Gade}[T]\equiv C \str^2(T\partial T^{-1})$, where $C$ is some small
constant and $S_{\rm T}$ is a contribution describing the coupling to
the leads.

The action $S[T]$ defines a one-dimensional representative of the
general family of 'chiral' non-linear $\sigma$-models.  In contrast to
its well investigated zero- and two-dimensional relatives -- much of
the results summarized in table~\ref{tab:1} have been derived within
these models -- the $1d$ variant has not been explored so far.

Much of our analysis of this model will rely on the standard analogy
between one-dimensional statistical field theory and zero-dimensional
quantum mechanics: $S[T]$ can be interpreted as the quantummechanical
action of a point particle propagating on the supersymmetric target
manifold, in our case $\GL$. The first term of the action represents a
kinetic energy, the third term a potential, and the second term,
linear in the 'velocity', coupling to a constant vector potential of
strength $(N+f)/2$. Quantum analogies of this type have been proven a
powerful technical tool in previous analyses of the standard
$\sigma$-models\cite{Ef:83}. However, the present case is special in
that the target manifold is so simple that an {\it intuitive}
interpretation of the quantum picture becomes straightforward.
Indeed, the fields $T$ have the explicit matrix representation $T =
\left(\matrix{u&\rho\cr\sigma&v}\right)$, where $\rho,\sigma (u,v)$
are Grassmann (commuting) variables, to be compared with the four or
eight dimensional matrices entering the standard $\sigma$-models.
Later on we will see that convergence criteria constrain the component
$u$ to be positive real, while $v=\exp(i\phi)$ is a pure phase. Thus,
temporarily leaving the Grassmann variables aside, our model describes
quantum propagation on a (half)line and on a circle, respectively.
Notice that the latter component is topologically non-trivial.

At this stage, the role of the vector potential contribution to the
action becomes evident. While inessential in the non-compact sector of
the theory, in the compact, circular sector, it desribes the presence
of an Aharonov-Bohm type magnetic flux. This analogy explains the
presence of phenomenona periodic in the channel number. An even number
of channels translates to an integer number of flux quanta through the
ring, which has no effect. However for $N$ odd or, alternatively, a
finite staggering parameter $f$, a fractional flux pierces the system
and this influences both, spectrum and dynamics of the quantum
system. To develop the picture somewhat further, notice that the
conductance, essentially the transition probability through the
system, maps onto the Green function of the quantum system evaluated at
imaginary 'time' $L$. Imagining the latter represented through a
spectral decomposition, the large $L$ behaviour depends on the low
lying portions of the quantum spectrum, in particular the discrete,
and flux periodic level structure of the compact sector of the theory.
Later on we will see that for half integer flux (i.e. $N$ odd) there
is a zero-energy level ($\rightarrow$ absence of localization) while
for all other magnetic configurations the spectrum is gapped
($\rightarrow$ exponential localization.) This mechanism, and its
relevance for the localization behaviour of the system was first
analysed by Martin Zirnbauer\cite{fn1}.  Similar but slightly more
elaborate arguments can be used to understand the profile of the DoS.

We finally mention some intriguing parallels to the physics of the
antiferromagnetic spin chain. According to Haldanes conjecture, a
chain of spins with half-integer (integer) $S$ is in a long range
ordered (disordered) phase\cite{haldane83}. It has also been found
(see Ref. \cite{affleck88} for review) that for a chain with staggered
hopping amplitude $j$, the strict integer/half-integer pattern is
violated, e.g. an integer chain can be fine-tuned to an ordered phase.
Technically, the system is described by a $\sigma$-model with a
topological term not dissimilar to the one above.  The differences are
that in the spin case (a) the base manifold is $1+1$-dimensional (a
{\it quantum} chain), (b) the target manifold is the two-sphere
$\simeq {\rm SU}(2)/{\rm U}(1)$, and (c) the topological term
classifies field configurations according to the number of coverings
of the sphere (instead of winding numbers around the circle, as in our
case). In the spin case, the coupling constant of the topological term
is given by $S+j$, i.e. spin replaces channel number and staggering
plays a similar role as in our case. (In fact, the linear coupling of
the topological term to the staggering amplitude follows,
independently of the model, from parity-conservation arguments to be
discussed below.) Beyond these apparent technical parallels, the
connection between the quenched disordered sublattice and the spin
chain, respectively, is not understood.

The rest of the paper is organized as follows. In section
\ref{sec:defin-model} we quantify the definition of the model. Its
field theory representation is introduced in section
\ref{sec:field-theory}. In section \ref{sec:transf-matr-appr} the
$\sigma$-model transfer matrix approach, i.e. the representation of
observables in terms of the quantum Green function is discussed. This
formulation is then applied to the calculation of conductance (section
\ref{sec:conductance}) and density of states (section
\ref{sec:density-states}). We conclude in section \ref{sec:summary}.

\section{Definition of the model}
\label{sec:defin-model}

We begin this section by upgrading the above qualitative introduction
of the sublattice system to a more quantitative formulation. Quantum
transport through the bulk of the system is described by the
Hamiltonian
\begin{equation}
  \label{eq:1}
  \hat H = \sum_{\langle ll'\rangle} \, c_{l,\mu}^\dagger
  \left(t_{ll'}\delta^{\mu\mu'}+ R^{\mu \mu'}_{ll'}\right) c_{l'\mu'},
\end{equation}
where $c_{l\mu}^\dagger$ creates a spinless electron at site $l$ in state
$\mu=1,\dots,N$, the sum extends over nearest neighbour sites, and
$R_{ll'}$ are $N\times N$ Gaussian distributed random hopping matrices  
with moments
\begin{eqnarray*}
&&\langle R_{ll'}^{\mu\nu} \rangle = 0,\\
&&\langle R_{ll'}^{\mu\nu} R_{ll'}^{\nu'\mu'}\rangle = {\lambda^2\over
  N}\delta_{\mu\nu} 
 \delta_{\mu'\nu'},\qquad \lambda \ll 1. 
\end{eqnarray*}
Apart from the Hermiticity condition $R_{ll'} = R^\dagger_{l'l}$
matrices on different links are statistically independent.  These
random matrices compete with the regular contribution to the
hopping matrix elements, $t_{ll'}$. To be specific, we set $t_{ll'}= 1
+(-) a$ if the smaller of the two neighbouring site indices $l$ and
$l'$ is even (odd). The real parameter $a$ is a measure for the
staggering strength.  Notice that $t_{ll} = {\cal O}(1)\gg \lambda$
implies that we are dealing with a weakly disordered system.

At both ends, a number of sites is coupled to leads (see figure
\ref{fig:figure1}). Quantum propagation within these leads is assumed
to be generic, i.e.  governed by a Hamiltonian without sublattice
symmetry.  To describe the coupling, we add to our bulk Hamiltonian a
tunnelling contribution
\begin{eqnarray*}
&&  \hat H_T = \hat H_T^L + \hat H_T^R,\\
&& \hat H_T^L = \sum_{l=1,2,\dots} \left(c_{\alpha}^\dagger
  W^L_{\alpha a} d^L_a +
{\rm h.c.}\right),\\
&& \hat H_T^R = \sum_{l=L,L-1,\dots} \left(c_{\alpha}^\dagger W^R_{\alpha
    a} d^R_a + 
{\rm h.c.}\right),
\end{eqnarray*}
where $d^{L(R)\dagger}_a$ creates an electron propagating in the left
(right) lead in a certain state $a=1,\dots, M\gg 1$ and we have
introduced a composite index $\alpha =(l,\mu)$ comprising site and
orbital index of the bulk system.  The coupling matrix elements
$W^{L/R}_{\alpha}$ are subject to the orthogonality
relation\cite{Hans:1}
\begin{eqnarray}
\label{eq:6}
&&\sum_{\mu} W^L_{a,l\mu}W^L_{l' \mu,b}= f(l) \delta_{ab}\delta_{ll'},
\nonumber\\ 
&&\sum_{\mu} W^R_{a,(L-l)\mu}W^R_{(L-l') \mu,b}= f(l)
\delta_{ab}\delta_{ll'},   
\end{eqnarray}
where $f(x)$ is some envelope function decaying on a scale of ${\cal
  O}(1)$ and normalized through $\sum_l f(l) = \gamma\ll 1$. The
function $f$ and parameter $\gamma$ describe profile and strength of
the coupling, respectively. Why did we introduce the multi-site
coupling operators (\ref{eq:6}) instead of connecting just the two
terminal sites $l=1,L$ to the lead continuum? Modelling the coupling
in a more general way is motivated by the presence of the site number
staggering phenomena mentioned in the introduction. The above
implementation of the coupling operator is sufficiently flexible to
selectively probe these effects (sections \ref{sec:cond-short-syst-1}
and \ref{sec:density-states-short}) {\it or} to average over any
boundary oscillatory structures (sections \ref{sec:reduction-problem}
and \ref{sec:heat-kernel-finite}.)

To conclude the definition of the problem, let us introduce the
Green function, 
\begin{equation}
  \label{eq:2}
 G(z) = \left(z- \hat H +  i\pi ({\rm sgn\, Im\,}z)\sum_{C=L,R}
   \hat W^C\hat W^{CT}\right)^{-1},
\end{equation}
where $\hat W^C\hat W^{CT} = \{ \sum_a W^C_{\mu l, a} W^C_{a,\mu' l}
\}$ is an operator describing the escape of electrons from the bulk
system into the leads\cite{Hans:1,Hans:2}. Expressed in terms of these
objects, the Landauer conductance of the system assumes the form
\begin{equation}
  \label{eq:5}
g = (2\pi)^2 \sum_{ab=1}^M
W^L_{a \alpha} W^R_{\beta b} W^R_{b \beta'} W^L_{\alpha' a}
\;\;\left\langle G_{\alpha \beta}(0^+) G_{\beta'\alpha'}(0^-) \right \rangle.
\end{equation}
Our second quantity of interest, the density of states per site, is
given by the standard expression
\begin{equation}
  \label{eq:7}
\nu(\epsilon) = -{1\over \pi L} {\rm \, Im \,}  \sum_\alpha
\langle G_{\alpha \alpha}(\epsilon^+)\rangle.
\end{equation}

Finally, to prepare the field theory formulation, let us consider the
symmetries of the problem. Expressed in the notation introduced above,
the chiral symmetry assumes the form $[\hat H,\hat \sigma_3]_+=0$,
where $\hat \sigma_3 = \{ (-)^l \delta_{ll'}\}$ and $\hat H$ denotes
the {\it bulk} part of the Hamiltonian. (Coupling to a non-sublattice
continuum breaks chirality.)  The presence of this symmetry implies
invariance under the two parameter family of transformations
\begin{eqnarray}
  \label{eq:42}
  c^\dagger_l \to c^\dagger_l e^{-z_1},\qquad& c_l \to e^{-z_2} c_l,&\qquad l
 \;\rm{even},\nonumber\\
c^\dagger_l \to c^\dagger_l e^{+z_2},\qquad& c_l \to e^{+z_1} c_l,&\qquad l
 \;\rm{odd},
\end{eqnarray}
where $z_{1,2}$ are complex numbers.  Depending on the choice of these
parameters, (\ref{eq:42}) expresses the standard ${\rm
  U}(1)$-invariance of a model with conserved charge ($z_1=-z_2$), or
the axial symmetry characteristic for chiral systems ($z_1=z_2$). (For
the time being we treat the transformation as purely formal i.e.
ignore the fact that for a general choice $z_{1,2}$ the transformed
operators are no longer mutually adjoint.)  We will come back to
discussing the role of these symmetries after the effective field
theory of the system has been introduced.

\section{Field Theory}
\label{sec:field-theory}

The construction of the low energy effective field theory of the
sublattice wire follows the standard route of deriving nonlinear
$\sigma$-models of disordered fermion systems\cite{Efetbook}, there
are no conceptually new elements involved. Referring to Appendices
\ref{sec:field-integr-form} and \ref{sec:deriv-field-theory} for
technical details of this derivation, we here discuss structure and
key features of the resulting model.

As shown in the Appendices, the field theory representation of
conductance and DoS is given by
\begin{equation}
  \label{eq:12}
  g= -\left({M\pi\gamma\over 2}\right)^2 \left\langle
  (T-T^{-1})_{12}(0)(T-T^{-1})_{21}(L) \right\rangle,
\end{equation}
and 
\begin{equation}
  \label{eq:11}
  \nu(\epsilon) = {\nu_0\over 2 L} {\rm \, Re \,}\int_0^L dr \left\langle
  T_{11}(r)+T^{-1}_{11}(r)\right\rangle, 
\end{equation}
respectively. Here, the bulk DoS, $ \nu_0 = {N\over 2\pi}$, $T(r)$ is
a field taking values in the supergroup ${\rm GL}(1|1)$ and the
continuum variable $r\in [0,L]$ replaces the site index.  The angular
brackets stand for functional averaging
$$
\langle \dots \rangle \equiv \int {\cal D}T e^{-S[T]}\; (\dots)
$$
over a functional integral with action $S\equiv \int_0^L dr {\cal L}$
and effective Lagrangian 
\begin{equation}
  \label{eq:9}
  {\cal L} \equiv {\cal L}_{\rm fl} + {\cal L}_{\rm top} + {\cal
    L}_{\rm T} + {\cal L}_z + 
{\cal L}_{\rm Gade}. 
\end{equation}
The individual contributions are given by
\begin{eqnarray}
  \label{eq:4}
  &&{\cal L}_{\rm top} = -{N+f\over 2} {\rm \,str\,}(T\partial_r
  T^{-1}),\nonumber\\ 
  &&{\cal L}_{\rm fl} = -{\xi\over 16} 
 \str (\partial_r T\partial_r T^{-1}),\nonumber \\
&&  {\cal L}_{z} = -i {z\pi\nu_0\over 2} {\,\rm
  str\,}(T+T^{-1}),\nonumber\\
&& {\cal L}_{\rm T} = {\pi M \gamma \over 2}{\,\rm \,str\,}(
T(r) + T^{-1}(r)) \left[\delta(r)+\delta(L-r)\right],\nonumber \\
&& {\cal L}_{\rm Gade} \equiv C \left[\str(T\partial_r
  T^{-1})\right]^2,
\end{eqnarray}
where $C$ is a coupling constant that need not be specified other than
that it is small, $C= {\cal O}(1)\ll (N,M)$ and the parameter
$f\equiv{2Na\over \lambda^2}$ measures the degree of staggering.
(Notice our $f$ is identical to the control parameter $f$ defined in
Ref.\cite{BMSA:98}.) Finally, we have introduced a parameter
$$
\xi \equiv {N\over 2\lambda^2}
$$
which will later identify as the localization length of the system.

To prepare the further discussion of the functional expectation
values, let us briefly discuss the internal structure of the field
theory.  We first note that save for the values of the coupling
constants, the structure of the action (\ref{eq:9}) can be anticipated
from inspection of Eq.  (\ref{eq:42}) and its supersymmetric
extension: the field theory approach starts out from a promotion of
the fermionic operator representation (\ref{eq:1}) to a supersymmetric
formulation in terms of Bose and Fermi fields. Within the latter
representation (cf. e.g. Eq.  (\ref{eq:3})), the space of permissible
invariance transformations is enlarged. The formerly complex
parameters $e^{z_i}$ are replaced by two-dimensional matrices $T_i$
acting on the bosonic and fermionic components of the theory.  Any
sensible subsequent manipulation done on the functional integral must
respect this invariance property.  On the level of the effective
theory described by $S[T]$, the transformation acts as $T\to T_1 T
T_2$ and indeed one verifies that the contributions ${\cal L}_{\rm
  top}, {\cal L}_{\rm fl}$ and ${\cal L}_{\phi}$ of (\ref{eq:9}) are
invariant under this operation. Further, the two building blocks
$\str(T\partial T^{-1})$ and $\str(\partial T\partial T^{-1})$ are the
{\it only} operators with $\le 2$ derivatives compatible with the
global ${\rm GL}(1|1)$ symmetry. In other words, the gross structure
of the bulk action readily follows from the invariance criterion.  For
finite energies or coupling to the leads chirality is broken and
global ${\rm U}(1)$ remains the only symmetry of the Hamiltonian.
Within the supersymmetry formulation, the set of allowed
transformations is then reduced down to configurations with
$T_1=T_2^{-1}$ (the super-generalization of ${\rm U}(1)$). The
operator $\str(T+T^{-1})$ is the minimal positive (see below) choice
compatible with the restricted symmetry. Summarizing, the three terms
$\str(T\partial T^{-1})$, $\str(\partial T\partial T^{-1})$, and
$\str(T+T^{-1})$ exhaust the set of operators compatible with the
global transformation behaviour of the model.

As for the coupling constants of the theory -- not specified by
symmetry arguments but all derived in Appendix
\ref{sec:deriv-field-theory} -- notice that contrary to the standard
$\sigma$-models of systems with WD symmetry two, instead of just one
second derivative operators appear in the action. {\it
  Mathematically}, the reason for the appearance of two contributions
is that the target manifold of the theory, $\GL$ is a reducible
symmetric space; it decomposes into two irreducible factors, a point
discussed in detail in Ref.  \cite{zirnbauer96}. Each of these factors
can be endowed with its own metric which implies the existence of two
independent second derivative operators in the theory. {\it
  Physically}, the presence of the non-standard operator $\sim
\str^2(T\partial T^{-1})$ has profound consequences for the behaviour
of the $2d$-version of the field theory\cite{gade93}: the RG analysis
of the model shows that the coupling constant of this operator grows
under renormalization while driving the coupling of the energy
operator $\sim \str (T+T^{-1})$ to large values. At the same time the
coupling constant of the standard gradient operator $\sim\str
(\partial T\partial T^{-1})$, essentially the conductance, remains
un-renormalized.  In one dimension, the situation is different.
Contrary to what one might expect, the contribution ${\cal L}_{\rm Gade}$ is not remotely as important as in two dimensions, and it is
another operator that drives the localization behaviour of the model.
In parentheses we remark that the target space of the transfer matrix
approach to the problem, ${\rm GL}(M)/{\rm U}(M) \simeq {\rm
  SL}(M)/{\rm SU}(M) \times {\cal R}^+$ factors into two components,
too\cite{brouwer00:_nonun}.  Accordingly, the Fokker-Planck equation
governing the 'Brownian motion' on that space is controlled by {\it
  two} independent coupling constants, both determined by the
microscopic definition of the model.  The second of these
contributions, essentially the analogue of our Lagrangian ${\cal
  L}_{\phi}$, leads to non-universal corrections to the overall
picture which perish in the limit of a large number of channels.

A second aspect discriminating the present model from its WD relatives
is the appearance of a first order gradient operator in the action. In
fact, the presence of this contribution might raise suspicion:
although allowed by symmetry, $\str(T\partial_r T^{-1})$ is not
invariant under space reflection $r\to -r$, in contrast to the
microscopic parent model (for $a=0$.)  The resolution of this puzzle
lies in the fact that $S_{\rm top}\equiv \int {\cal L}$ is of
topologcial origin and, although not manifestly so, {\it does} respect
the space inversion property. We will discuss this point momentarily
after the internal structure of the field manifold and the integration
measure have been specified.

Functional integrals can only be defined on manifolds that are
Riemannian, i.e. endowed with a positive metric. The supergroup ${\rm
  GL}(1|1)$ (like all other supersymmetric spaces that appear in the
context of field theories of disordered Fermi systems) does not fulfil
this criterion, a point discussed in detail in Ref.
\cite{zirnbauer96}. However, it does contain a maximally Riemannian
{\it sub}manifold ${\cal M}$, defined as follows: the boson-boson
block of ${\cal M}$, a one-dimensional manifold by itself, is
isomorphic to the non-compact symmetric space ${\rm GL}(1)/{\rm
  U}(1)\simeq {\cal R}^+$, i.e. the positive real numbers. (This space
is trivially Riemannian.) The fermion-fermion block is isomorphic to
the compact symmetric space ${\rm U}(1)$. No limitations in the
Grassmann valued boson-fermion sectors of the theory are needed since
the whole issue of convergence does not arise here.  Summarizing,
${\cal M} = {\cal R}^{+} \times {\rm U}(1)$, where the notation is
symbolic, specifying the boson-boson and fermion-fermion sector,
respectively.

A convenient field representation respecting these  convergence
criteria is given by
\begin{eqnarray}
\label{eq:8}
  T = k a k^{-1},\qquad
k = \exp\left(\matrix{&\eta\cr\nu&}\right),\qquad
a = \exp\left(\matrix{x &\cr&iy}\right),
\qquad x,y\in {\cal R}.
\end{eqnarray}
In this parameterization, the group integration $\int dT$ extends over
the degrees of freedom $x,y,\eta,\nu$, without further constraints.
The invariant group measure associated to the representation
(\ref{eq:8}) are defined in Eqs. (\ref{eq:22}) and (\ref{eq:28}).

We are now in a position to discuss the role of the contribution
$S_{\rm top}$. First notice that for sufficiently strong lead coupling
the boundary action ${\cal L}_{\rm T}$ projects the fields onto the
group origin $T(x,y)=\openone$, i.e.  enforces $x(0)=x(L)=0$ and
$y(0)=2\pi k$, $y(L)=2\pi k'$, where $k,k'$ are integer.  As a
consequence, the first derivative operator can be written as
 \begin{eqnarray}
\label{eq:35}
&&S_{\rm top}[T] = -{N+f\over 2} \int dr {\rm \,str\,}(T\partial_r
  T^{-1})=\nonumber\\
&&\qquad= {N+f\over 2} \int dr \partial_r {\rm \,str\,}\ln T=\nonumber\\ 
&&\qquad= {N+f\over 2} {\rm \,str\,}(\ln T(L)-\ln T(0))=\nonumber\\ 
&&\qquad= i \pi  (N+f)(k-k').
 \end{eqnarray}
This makes the topological nature of the term manifest: it counts 
winding numbers in the fermionic sector of the theory. The integer
$k-k'$ is a topological invariant characterizing each field
configuration $T(r)$. Further notice that for $f=0$, 
$$
e^{-S_{\rm top}[T]} = e^{+i\pi N (k-k')} = e^{-i\pi N(k-k')}.
$$
Since it is the exponentiated action and not the action itself that
matters, the last identity tells us that the global {\it sign} of the
first gradient operator is irrelevant (all for $f=0$). This settles
the above raised issue of the behaviour of the model under space
reflection: although the first order derivative is not invariant under
$r \to -r$ the exponentiated action is. 

\section{Transfer Matrix Approach}
\label{sec:transf-matr-appr}

Principal aspects of the system properties we are going to analyse are
non-perturbative, i.e. cannot be obtained as power series in the
coupling constants of the action.  Progress with such type of problems
can be made by applying the $\sigma$-model transfer-matrix
technique\cite{Ef:83}, an approach conceptually similar to the DMPK
formalism.

We begin by defining  the two functions
\begin{eqnarray*}
&&Y_L(T_1,T_2,r) \equiv \int_{T(0)=T_1\atop T(r)=T_2} {\cal D}T e^{-\int_0^r dr {\cal
    L}[T]},\\  
&&Y_R(T_1,T_2,r) \equiv \int_{T(r)=T_1\atop T(L)=T_2} {\cal D}T
e^{-\int_r^L dr {\cal L}[T]}.
\end{eqnarray*}
Expressed in terms of these objects, the DoS assumes the form
\begin{eqnarray}
  \label{eq:13} 
\nu(\epsilon) = {\nu_0 \over 2 L} {\,\rm Re\,} \int_0^L
dr 
\int dT \; Y_L(\openone,T,r)\; (T_{11}+T^{-1}_{11})\; Y_R(T,\openone,L-r),
\end{eqnarray}
where we have used that for sufficiently strong coupling to the leads,
the boundary configurations $T$ are close to unity. (Throughout much
of this paper we will consider the DoS of coupled systems.  For
sufficiently large systems, the choice of boundary conditions is
inessential, a point to be verified below.)  Similarly, the
conductance obtains as
\begin{eqnarray}
\label{eq:14}
&&    g= -\left({\pi M\gamma\over 2}\right)^2 \int dT \;
dT'\;\times\nonumber\\  
&&\hspace{2.0cm}\times e^{-{\pi M
        \gamma \over 2} \str(T+T^{-1})}
  (T-T^{-1})_{12}\; Y_L(T,T',L)\; (T'-T'^{-1})_{21} e^{-{\pi M
        \gamma \over 2} \str(T'+T'^{-1})}.
\end{eqnarray}
From Eqs. (\ref{eq:13}) and (\ref{eq:14}) it is clear that the
functions $Y_{L,R}$ encode the essential system properties we are
interested in.

As a first step towards the computation of these objects let us
explore how the symmetries of the action translate to symmetries of
$Y_{L,R}$.  We first consider the case $z=0$, relevant to the analysis
of the conductance, where the action is fully invariant under
$\GL$-transformations. The invariance ${\cal L}[T]={\cal L}[T_1 T
T_2]$, $T_{1,2}={\rm const.}$ then directly implies
$Y_{L,R}(T,T',r)=Y_{L,R}((T_1T T_2),(T_1 T' T_2),r)$. From this
identity one readily verifies that
\begin{eqnarray*}
&&Y_R[T,T',r]=Y_R[T T'^{-1},\openone,r]=
Y_R[T^{'-1} T,\openone,r],\\
&&Y_L[T,T',r]=Y_L[\openone,T' T^{-1},r]=
Y_L[\openone,T^{-1} T',r],\qquad(z=0).  
\end{eqnarray*}
In other words, for $z=0$ the arguments of the heat kernels enter in
an invariant product type form and it suffices to consider the reduced
functions 
\begin{eqnarray}
  \label{eq:16}
Y_{R}(T,r)\equiv Y_{R}(T,\openone,r),\qquad  Y_{L}(T,r)\equiv
Y_{L}(\openone,T,r),
\end{eqnarray}
depending on a single argument only. The same invariance property (now
evaluated for $T_2=T_1^{-1}$) implies that $Y_{L,R}(T,r) = Y_{L,R}(T_1
T T_1^{-1},r)$. Imagining the argument $T$ to be represented in the
polar decomposition (\ref{eq:8}) and setting $T_1=k^{-1}$ the argument
can be reduced to the diagonal matrix of eigenvalues $a$:
\begin{eqnarray}
  \label{eq:15}
Y_{L,R}(kak^{-1},r)=Y_{L,R}(a,r).  
\end{eqnarray}
For $z\not=0$, the invariance of the theory collapses down to
transformations with $T_2=T_1^{-1}$. However, the representation of
the DoS above implies that we are solely interested in functions of
the type (\ref{eq:16}), with second argument set to unity, anyway.
Since transformations $T\to T_1 T T_1^{-1}$ are still permissible,
these objects depend on the eigenvalues of the argument matrix only,
as before for the case $z=0$.  Summarizing, in the analysis of both
conductance and DoS, it is sufficient to consider functions $Y_{L,R}$
depending on a single argument with  invariance property (\ref{eq:15}).

Following the basic philosophy of the transfer matrix approach, we
will compute the functions $Y_{L,R}(T,r)$ iteratively, by asking how
much they change under infinitesimal variation of the arguments $r\to
r+\epsilon$. Considering the function $Y_L$ for definiteness, we first
notice that, by definition,
$$
Y_L(T,r+\epsilon)=\int {\cal D} T e^{-\int_r^{r+\epsilon}dr' {\cal
    L}[T]} Y_L(T(r),r).
$$
For sufficiently small $\epsilon$, the action can be expanded and we obtain
\begin{eqnarray*}
&&  Y_L(T,r+\epsilon)-Y_L(T,r) = \int dT' e^{-{\xi\over 16 \epsilon}
  \str(TT'^{-1} + T^{-1}T')}\times\\
&&\hspace{3.0cm}\times  
e^{  {N+f\over 2} \str (T'T^{-1}) + i {z\pi\nu\epsilon \over 2} \str(T+T^{-1})}
\left(Y_L(T',r)-Y_L(T,r)\right),
\end{eqnarray*}
where we have used that, due to supersymmetry, $ \int {\cal D}T
\exp(-\int_r^{r+\epsilon} {\cal L}[T]) \times 1 = 1$. We have also
temporarily set the coupling constant of the operator ${\cal L}_{\rm Gade}$
to zero. As mentioned above, this term does not have much relevance in
the present context. We will briefly discuss its role in section
\ref{sec:role-gade-term}.

From hereon, the derivation of an evolution equation for $Y_L$ is
conceptually straightforward: the exponential weight factor in the
first line of the equation enforces that $T'$ is close to $T$,
symbolically, $T'T^{-1} = \openone + {\cal O}(\epsilon)$. We should
thus expand $T'$ around $T$ and evaluate the integral $\int dT'$
perturbatively in $\epsilon$. This expansion is most conveniently done
in the polar coordinates introduced above (because the heat kernel
depends on the radial degrees of freedom, only.) As a result of a
calculation similar but much more simple than the one for the standard
$\sigma$-models with their larger matrix fields\cite{Efetbook} one
obtains the Schr\"odinger type equation
\begin{eqnarray}
  \label{eq:17}
&&  \left(\mp \partial_t - 4(\D \pm \A)^2 + V(x,y)
\right)Y_{L\atop R}(a,r)=0,
\end{eqnarray}
where $ V(x,y)= -is (\cosh(x)-\cos(y)) $ and we have introduced the
dimensionless parameters
$$
t \equiv {r\over \xi},\qquad s= \pi \nu\xi \epsilon.
$$
Physically, $t$ is the length coordinate measured in units of the
localization length $\xi$ and $s$ the energy measured in units of the
single particle level spacing $\Delta_\xi \equiv \xi \nu$ of a system
of length $\xi$. 
Further, the
symbol $\D=(D_x,D_y)^T$ denotes a vector differential operator defined
through
\begin{eqnarray}
\label{eq:21}
&&D_{x} = \partial_{x} - {1\over 2}
\coth\left({x-iy\over 2}\right),  \nonumber\\
&&D_{y} = \partial_{y} + {i\over 2}
\coth\left({x-iy\over 2}\right),
\end{eqnarray}
where the constant vector
\begin{equation}
  \label{eq:24}
  \A = {N+f\over 2}(1,-i)^T.
\end{equation}
Finally, evaluation of (\ref{eq:17}) for asymptotically small times $t
\searrow 0$ leads to the initial condition
\begin{equation}
  \label{eq:18}
  \lim_{t\to 0} Y_{L,R}(x,y,t)=\delta(x,y) \equiv \lim_{t\to 0}e^{-{1
      \over 8t} (x^2 + y^2)}.
\end{equation}

It is very instructive to interpret the structure of the evolution
equation (\ref{eq:17}) in the light of the analogy between field
theory and point particle quantum mechanics on $\GL$ discussed in the
introduction.  Within this picture, the functions $Y_L[T,r]$ acquire
the meaning of {\it Green functions}, i.e. transition amplitudes for
the propagation between the origin of the manifold $T=\openone$ at
time $0$ and a final configuration $T$ at time $t$.  The evolution
equations (\ref{eq:17}) becomes a a time-dependent Schr\"odinger
equation with quantum Hamiltonian, $H = -2 (\D-\A)^2 + V(x,y)$. While
the term $V(x,y) = -is (\cosh(x)-\cos(y)) = -{is\over 2}
\str(T+T^{-1})$ simply represents the potential inherited from the
Lagrangian, the 'kinetic' operator is more interesting.  The covariant
structure $(\D-\A)^2$ describes minimal coupling to the constant
vector potential where the unfamiliarly looking structure of the
derivative operator $\D$ is a consequence of the fact that our
particle lives on a curved manifold. Indeed, it is straightforward to
verify that for $\A=0$
$$
\D\cdot \D = \sum_{i=x,y} J^{-1}\partial_i J\partial_i,
$$
where $J(x,y)$, specified in Eq.~(\ref{eq:28}), is the square root
of the determinant of the metric tensor on $\GL$. This structure
identifies $\D \cdot \D$ as the radial part of the Laplace operator on
$\GL$. ('Radial', because the two coordinates $x$ and $y$, spanning a
maximally commutative sub-algebra of the Lie algebra of $\GL$.)

Summarizing, we have identified $H$ as the Hamiltonian of a charged
particle on the group manifold $\GL$. Our next task will be to compute
its Green functions $Y_{L/R}$. We begin by considering the case of a
free particle, $V=0\leadsto \epsilon=0$. The solution of this problem
will contain the information needed to compute the conductance.

\section{Conductance}
\label{sec:conductance}

In this section the general formalism developed above is employed to
compute the conductance $g$ of the system. Our main objective will be
to understand the impact of topology on the localization behaviour of
the system. However, before embarking on this analysis it
is tempting to digress for a moment and to briefly consider the
transport behaviour of {\it short} systems, specifically the
aforementioned anomalous sensitivity to the coupling to the leads.
Being not directly related to the mainstream of the paper, the
technicalities of this discussion have been deferred to Appendix
\ref{sec:cond-short-syst} and we here restrict ourselves to a summary
of results.

\subsection{Digression: Conductance of Short Wires}
\label{sec:cond-short-syst-1}

In the present context, the phrase 'short' means that systems in the
quantum dot regime $L<\xi/(M\gamma)$ are considered: the conductance is
not so much determined by the bulk transport properties of the system
but rather by the strength of the coupling to the leads.  Moreover,
and this is a speciality of the sublattice system, the {\it parity} of
the coupling, i.e. the even- or oddness of the connecting sites, turns
out to play a crucial role. More precisely we find that (a) for
systems where at both ends a number of sites of alternating parity are
connected -- the setup considered in much of this paper -- the short
system conductance is given by
$$
g= {M\pi \gamma\over 2},
$$
in accord with the behaviour of conventional quantum dots. The same
result obtains (b) for systems where only one site at either end is
coupled and these sites have opposite parity (even/odd or odd/even).
However, (c) for single site coupling with even/even or odd/odd
connectors, the conductance {\it vanishes}.  To heuristically
understand the phenomenon, it is instructive to consider the toy-model
case of a strictly one-dimensional clean sublattice system. For zero
energy, the wave length of current carrying excitations is
commensurable with the lattice spacing. This means that the relevant
quantum wave functions have nodes at alternating sites. E.g., for an
even-even configuration a state entering from the right has zero
quantum amplitude at the exit site on the right. This implies a total
blockade of electric current.  A less obvious fact is that this
phenomenon survives generalization to many channels and disorder.

We repeat that all these results are obtained for short systems; field
fluctuations, describing the propagation of spatially non-uniform
diffusive excitations, are neglected. An interesting question, not
considered in this paper, is how such modes would affect the
conductance as the system size is increased.

\subsection{Reduction of the Problem to (0-dimensional) quantum
  mechanics on $\GL$}
\label{sec:reduction-problem}

We next turn back to the discussion of large systems (equilibrated
coupling of type (a) understood.) Inspection of the basic expression
(\ref{eq:14}) shows that the problem factorizes into doing the
boundary integrals and analysing the bulk behaviour of the heat kernel,
respectively.

We begin by discussing the boundary regions.  Following a line of
arguments developed in Ref.\cite{mirlin94}, we first notice that due
to the exponential weights $\sim \exp(-{\pi M \gamma\over
  2}\str(T+T^{-1}))$, the integrands are confined to the immediate
vicinity of the group origin.  This suggests to write $T=\exp W$, and
do the integrals over generators $W$ in a Gaussian approximation.
Setting $W=\left(\matrix{u&\sigma \cr \rho & iv}\right)$ and expanding
in coefficients we arrive at integrals of the structure,
\begin{eqnarray*}
g={\rm const.}\times  \int du dv d\rho d\sigma e^{-{\pi M \gamma \over
    4}(u^2 + v^2+ 2 \sigma 
  \rho)} \sigma F(T(u,v,\sigma,\rho)),   
\end{eqnarray*}
where the symbol $F(T)$, represents the functional dependence of the
heat kernel on the boundary field.  Evaluation of the Gaussian
superintegral leads to 
$$
g = {\rm const.} \times {1\over \pi M \gamma} \partial_\rho
F(T)\big|_{T=\openone}. 
$$
Doing the same procedure for the second boundary integral and
fixing factors we obtain
$$
g = \partial_\rho \partial_{\sigma'}Y_L(T,T',L)\big|_{T=T'=\openone}=
\partial_\rho \partial_{\sigma'}Y_L(TT'^{-1},L)\big|_{T=T'=\openone},
$$
where the second expression contains the one-argument heat kernel
introduced in Eq. (\ref{eq:16}). According to this expression, the
conductance is obtained by second order expansion of the heat kernel
around the origin. We next note that due to the invariance property
(\ref{eq:15}) the expansion starts as
\begin{equation}
  \label{eq:23}
Y_L(\tilde T=e^{\tilde W},L) = 1 + c_1 \str(\tilde W) + c_2 \str(\tilde W^2)
+ c_3 [\str(\tilde W)]^2+\dots  .
\end{equation}
Now, in our case,
$$
\tilde W = \ln(TT'^{-1})= \ln\left(\exp\left(
    \matrix{&0\cr\rho&}\right) \exp\left( \matrix{&-\sigma'\cr
      0&}\right) \right)=\left(\matrix{{1\over 2}\sigma'\rho
    &-\sigma'\cr \rho &{1\over 2}\sigma'\rho}\right).
$$
Substitution of this expression into (\ref{eq:23}) and
differentiation leads to
\begin{equation}
  \label{eq:34}
  g=2c_2,
\end{equation}
i.e.  the problem has been reduced to fixing the coefficients of the
series expansion of $Y_L$.  Notice that this series representation is
totally determined by the features of the bulk system; all aspects
related to the coupling to the leads have disappeared from the
problem.

We now have to face up to the principal task, the calculation of the
heat kernel. Its interpretation as the Green function of the problem
suggests to begin by representing $Y_L$ through a formal spectral
decomposition: consider Eq.~(\ref{eq:17}) at zero potential, $
\left(-\partial_t - 4 (\D-\A)^2 \right)Y_{L}(x,y,t)=0$, and suppose we
had managed to find a set of eigenfunctions,
\begin{equation}
  \label{eq:26}
  -4(\D-\A)^2 \Psi^{(r)}_n = \epsilon_n \Psi^{(r)}_n.
\end{equation}
Assuming completeness, we can  then span the heat kernel as
$$
Y_L(x,y,t) = \sum_n c_n \Psi^{(r)}_n(x,y) e^{-\epsilon_n t},
$$
where the expansion coefficients are determined through the initial
condition (\ref{eq:18}). Provided the spectrum is suitably structured
(positive and gapped against some low lying levels), and keeping in
mind that we are interested in asymptotically large values of $t$, the
sum may be restricted to a limited set of $n$'s. The problem thus
reduces to (a) exploring the low-lying spectral content of the
operator $\D-\A$, and (b) fixing the expansion coefficients.

\subsection{Spectrum and Eigenfunctions of the 'Kinetic Energy
  Operator' on $\GL$}
\label{sec:spectr-eigenf-kinet}

In spite of the unfamiliarly looking coordinate representation of
$(\D-\A)^2$ analytic progress with the problem is straightforward, the
reason being that this operator is nothing but a plane wave operator
in disguise. To make the hidden simplicity of the problem manifest,
let us first remove the dependence of the differential operator on the
(pure gauge) potential $\A$: transformation $\Psi_n(x,y) \to
e^{{N+f\over 2}(x-iy)} \Psi_n(x,y) \equiv \hat \Psi_n(x,y)$ brings the
eigenvalue equation (\ref{eq:26}) into the form
$$
-4 \Delta \hat \Psi_n = \epsilon_n \hat \Psi_n,
$$
where $\Delta=\D\D$ is the radial part of the Laplacian on $\GL$.
Notice that the structure of this equation does {\it not} imply that
the vector potential has disappeared from the problem.  It has merely been
transferred from the differential operator to the boundary conditions
attached to the differential equation. While irrelevant in the
non-compact sector of the theory, changes of the boundary conditions
in the compact sector generally cause qualitative effects.  To
appreciate this point, notice that the un-gauged Hilbert space of the
problem has periodic boundary conditions, $ \Psi_n(x,y) =
\Psi_n(x,y+2\pi).  $ Yet, after the gauge transformation, $ \hat
\Psi_n(x,y) = \hat \Psi_n(x,y+2\pi) (-)^{N+f} $, i.e.  for $N$ odd or
non-zero staggering a transmutation to twisted boundary conditions has
taken place. Needless to say that this change bears consequences for
the spectral structure of the problem and, therefore, for the
transport behaviour of the system.

To make further progress, we subject the eigenvalue problem to the
similarity transformation
\begin{eqnarray*}
&&\hat \Psi \to  J^{1/2} \hat \Psi \equiv \tilde
\Psi,\nonumber\\
&&\Delta \to J^{1/2} \Delta J^{-1/2} \equiv \tilde \Delta =
\partial_x^2 + \partial_y^2.
\end{eqnarray*}
This change of representation entails an enormous simplification of
the problem. The uninvitingly looking operator $\Delta$ has become a
flat two-dimensional Laplace operator\cite{fn2}. However,
notice that the transformation by $J^{1/2}$ effects yet another change
in the boundary conditions: due to $J^{1/2}(x,0) = -J^{1/2}(x,2\pi)$,
the transformed states are subject to the condition
$$
\tilde\Psi_n(x,y) = \tilde \Psi_n(x,y+2\pi) (-)^{N+f+1}.
$$
At this point, the solution of the eigenvalue problem has become a
triviality. The equation
$$
-4(\partial_x^2 + \partial_y^2)\tilde \Psi_n(x,y) =
\epsilon_n \tilde \Psi_n(x,y),
$$
is solved by the exponentials $ \tilde \Psi_{k,l}(x,y) =
e^{ip_l x + ip_k y}$, where $(k,l)\equiv n$ are two 'quantum numbers',
$p_{k,l}$ the associated momenta and the eigenvalues $\epsilon_{kl} =
4(p_k^2 + p_l^2)$. From these states we obtain our un-gauged
and un-transformed original wave functions as
\begin{equation}
  \label{eq:25}
\Psi_{kl}(x,y) = \sinh\left({x-iy\over
    2}\right)  e^{(ip_l -{N+f\over 2}) x + i(p_k + {N+f\over 2}) y}.  
\end{equation}
To give this set of solutions some meaning, we need to specify the
range of permissible $k$'s and $l$'s.  In the compact sector the
situation is clear -- the circular boundary condition specified above
enforces $p_k= k- {N+f\over 2}$, with {\it half}integer $k$.  The
conditions to be imposed in the non-compact sector are tightly linked
to the integrability properties of our wave functions:

The space of radial functions on $\GL$ is endowed with a natural
scalar product, viz.
\begin{equation}
  \label{eq:30}
\langle f,g\rangle \equiv \int_{-\infty}^\infty dx \int_0^{2\pi} dy
J(x,y) f(x,y) g(x,y).  
\end{equation}
We demand that the eigenfunctions contributing to the spectral
decomposition of the heat kernel be square integrable w.r.t. $\langle
\;,\;\rangle$. Inspection of Eq. (\ref{eq:25}) shows that this
requirement enforces $p_l = l -i {N+f\over 2}$, where $l$ may be
arbitrary and real. 


Finally, we need to specify a set of functions sufficiently complete
to generate an expansion of the heat kernel. The present problem does
not come with a natural Hermitian or symmetric structure (i.e. for
finite $\A$ the kinetic energy neither symmetric nor
Hermitian.) However, defining a 'fake complex conjugation' through $
\bar \Psi_{kl} = \Psi_{-k,-l} $ it is straightforward to show that
\begin{equation}
  \label{eq:31}
\langle \bar \Psi_{kl},\Psi_{k'l'} \rangle = (2\pi)^2 \delta(l-l')
\delta_{kk'}.  
\end{equation}
We may thus attempt to represent sufficiently well behaved (for the
cautious formulation, see below) functions as
\begin{equation}
  \label{eq:32}
g(x,y) = \sum_k \int dl\; g_{kl} \Psi_{kl}(x,y),  \qquad
g_{kl} = (2\pi)^{-2} \langle \bar \Psi_{kl},g \rangle.
\end{equation}
Before applying this procedure to the heat kernel, let us summarize
our main findings for clarity: the radial Laplacian on $\GL$ is
diagonalized by the set of functions
\begin{equation}
  \label{eq:33}
\Psi_{kl}(x,y) = \sinh\left({x-iy\over
    2}\right)  e^{i(lx + ky)}.  
\end{equation}
where $k\in {\cal Z}+1/2$, $l$ real and the eigenvalues
are given by
\begin{equation}
  \label{eq:29}
  \epsilon_{kl} = 4 \left[ \left(k - {N+f\over 2}\right)^2 + \left(l -
      i {N+f\over 2}\right)^2\right].
\end{equation}
(Notice that the appearance of an imaginary part proportional to the
strength of the vector potential is due to the fact that for finite
${\bf A}$, the Hamiltonian of the theory is neither symmetric nor
Hermitian.)  The spectral decomposition of radial functions is defined
through Eqs.  (\ref{eq:30}), (\ref{eq:31}) and (\ref{eq:32}).

\subsection{Computation of the Conductance}
\label{sec:comp-cond}

We now apply the machinery developed in the last section to the
analysis of the heat kernel. In principle, the strategy seems to be
prescribed by what was said above. We should determine the Fourier
coefficients of the initial configuration (\ref{eq:18}), $\delta_{kl}
= \langle \bar \Psi_{lk},\delta\rangle$, from where the heat kernel
would follow as $Y_L(x,y,t) = (2\pi)^{-2} \sum_k \int dl\; \delta_{kl}
e^{-\epsilon_{kl} t} \Psi_{kl}(x,y)$. There is a problem with this
procedure, viz. the expansion coefficients of the
'$\delta$-distribution' $\delta(x,y)$ vanish. Indeed one verifies that
in the limit $t\to 0$, the support of the Gaussian in (\ref{eq:18})
shrinks to zero while its maximum remains limited by one. This readily
implies $\langle \Psi,\delta\rangle=0$. The reason for this
pathological behaviour is that our radial theory memorizes that it
derived from a supersymmetric parent theory. In a sense, supersymmetry
can be interpreted as a theory on a zero-dimensional background, i.e.
there is no singular 'volume factor' compensating for the vanishing
support of the $\delta$-function, as would be the case in spaces with
positive dimension.

The problem can be circumvented by a cute trick\cite{mirlin94}.
Instead of Fourier expanding $Y_L$, we consider the function $Y_L -1$.
Since unity by itself solves the heat equation, no harm has been done
and all that has changed is the boundary condition: $\lim_{t\to 0}
(Y_{L}(x,y,t)-1)= \delta(x,y)-1$, a function that equals minus unity
almost everywhere save for the origin where it vanishes.

We thus represent the heat kernel as 
\begin{equation}
  \label{eq:39}
Y_L(x,y,r) = 1-  (2\pi)^{-2} \sum_k \int dl \, 1_{kl} \Psi_{kl}(x,y)
e^{-\epsilon_n t},  
\end{equation}
where 
$$
1_{kl} = \langle \bar \Psi_{kl},1 \rangle = \int dx \int dy 
{e^{i(lx + ky)}\over \sinh\left({x-iy \over 2}\right)}=
{4\pi i \over l-ik}. 
$$
(To obtain the last equality, it is convenient to first do the
$x$-integral. Closure of the integration contour in the upper/lower
complex half plane for positive/negative $l$ yields a semi-infinite
sum over residues of the $\sinh$-function, along with a $y$-integral
that is of simple plane wave type. Doing sum and integral one obtains
the result.)

Substitution of this result into the expansion of the heat kernel now yields
\begin{eqnarray*}
&&  Y_L(x,y,t)-1 =  -{i\over \pi} \sum_k \int dl \, {1\over l-ik} \sinh\left({x-iy\over
    2}\right)  e^{i(lx + ky)}
e^{-\epsilon_{kl} t}\stackrel{{\cal O}(x,y)^2}{\longrightarrow}\\
&&\qquad \stackrel{{\cal O}(x,y)^2}{\longrightarrow}
{1\over 4\pi}\sum_k \int dl  \left(
{l+ik\over l-ik}(x-iy)^2 +  (x^2+y^2)\right)e^{-\epsilon_{kl} t}=\\
&&\qquad = {1\over 4\pi }\sum_k \int dl  \left(
{l+ik\over l-ik}(\str(W))^2 +  \str(W^2)\right)e^{-\epsilon_{kl} t},
\end{eqnarray*}
where in the last line we have switched back to a coordinate invariant
representation. Comparison with Eqs. (\ref{eq:23}) and (\ref{eq:34})
finally leads to the identification
\begin{eqnarray*}
&&g = {1\over 2\pi}\sum_{k\in {\cal Z}+1/2} \int_{-\infty}^{\infty} dl
e^{- {4L\over \xi}\left[ \left(k - {N+f\over 2}\right)^2 + \left(l -
      i {N+f\over 2}\right)^2\right]}=\\
&&\qquad  = {1\over 2}\left({\xi \over 4\pi L}\right)^{1/2}
\sum_{k\in {\cal Z}+1/2} 
e^{- {4L\over \xi} \left(k - {N+f\over 2}\right)^2}.
\end{eqnarray*}
where we have inserted the explicit form of the eigenvalues.
Notice that all manipulations leading from the original $\sigma$-model
representation to the above Gaussian integral representation have been
exact.

We next evaluate this result in the two limiting cases of physical
interest, $L\ll \xi$ (Ohmic regime) and $L\gg \xi$  (localized
regime.) Beginning with the Ohmic case, we first notice that for
$L \ll \xi$ many terms contribute to the $k$ summation implying that the
sum can be approximated by an integral. Thus,
\begin{eqnarray*}
&&  g \stackrel{L\ll \xi}{\approx} 
{1\over 2}\left({\xi \over  4\pi L}\right)^{1/2}
\int dk 
e^{- {4L\over \xi} \left(k - {N-f\over 2}\right)^2} = {\xi\over 16L}.
\end{eqnarray*}
As for any ordinary conductor, $g$ is inversely proportional to the
system size; the parity of the channel number does not play a role.

In the opposite case, $L \gg \xi$, only those discrete indices that
minimize the exponent contribute to the sum. Specifically, for an even
channel number and no staggering,
$$
g \stackrel{L\gg \xi}{\approx} 
\left({\xi \over  4\pi L}\right)^{1/2}
e^{- {L\over \xi}}, \qquad N\; \mbox{even}, f=0.
$$
In contrast, for $N$ odd and $f$ still zero, the exponent vanishes for
the half integer $k={N\over 2}$ and
$$
g \stackrel{L\gg \xi}{\approx} 
{1\over 2}\left({\xi \over  4\pi L}\right)^{1/2}, \qquad N\;
\mbox{odd}, f=0
$$
depends algebraically on the system size. Finally, it is clear that
for non-vanishing staggering, $f\not=0$, intermediate types of
behaviour are realized. E.g.,  an $N$ even chain with staggering
$f=\pm 1$ behaves like a symmetric $N$ odd chain, etc.

\subsection{The Role of the Gade Term}
\label{sec:role-gade-term}

Before leaving this section let us briefly discuss the role of the, so
far neglected, Gade operator $S_{\rm Gade}[T]$. The inclusion of this
term in the derivation of the heat equation is straightforward. As a
result, the planar Laplacian $\tilde \Delta = \partial_x^2 +
\partial_y^2$ gets replaced by
$$
\tilde \Delta =
\partial_x^2 + \partial_y^2 + \tilde \eta (\partial_x - i\partial_y)^2,
$$
where $\tilde \eta \equiv {16 C \over \xi}\propto N^{-1}\ll 1$.
This operator is still diagonalized by the plane waves discussed
above. The eigenvalues change to
$$
\epsilon_{kl}=4 \left[(1-\tilde \eta)\left(k - {N+f\over 2}\right)^2 +
  (1+\tilde \eta)\left(l - i {N+f\over 2}\right)^2\right].
$$
Recapitulating the computation of the conductance, one finds
that the small dilatation introduced by finite values of $\tilde \eta$ does
not affect the long range transport behaviour of the system. 

All this is compatible with the structure of the DMPK transfer matrix
approach to the problem. As mentioned above, the DMPK evolution
equation is controlled by {\it two} coupling constants. One of these,
in Ref.  \cite{brouwer00:_nonun} denoted by $\eta$, is small,
$\eta\sim {\cal O}(M^{-1})$ and becomes inessential in the limit of a
large number of channels. The analysis above suggests that the
coupling constant of the Gade operator, $\propto \tilde \eta$, and the
$\eta$ of the DMPK approach play the same role.  This analogy is
supported by the above discussed geometric structure of the target
spaces of the two theories.

\section{Density of States}
\label{sec:density-states}

We now turn our attention to the low energy density of states of the
sublattice system.  As with the conductance, we will first consider
the behaviour of short wires, and then discuss the localized regime.

\subsection{Density of states of short wires}
\label{sec:density-states-short}

As in section \ref{sec:cond-short-syst-1} we consider a short
sublattice wire of length $L<\xi /(M\gamma)$. The spectrum of such
systems exhibits structure on the scale of the mean level spacing. In
order not to blur these fine structures, the coupling to the external
leads will be switched off throughout this section.  Following the
logics of section \ref{sec:cond-short-syst-1}, one would then conclude
that the action reduces to $S[T]=\int dr {\cal L}_z[T] =
-i{s\over 2}\str(T+T^{-1})$, where the matrix $T$ parameterizes a zero mode
configuration, $s={\pi \nu \over \Delta}$ and $\Delta = (\nu L)^{-1}$
is the level spacing of the isolated system. (Temporarily deviating
from the convention of the rest of the paper, in this section $s$
measures the energy in units of the total level spacing and not the
level spacing of a localization volume.) This presumption is almost
but not quite correct.  The point is that our so far discussion of the
low energy action implicitly assumed that the number of sites of the
system is even. In the opposite case, an extra contribution $S_{{\rm
    top},2}[T] = N \strln(T(L))$, derived and discussed in section
\ref{sec:goldst-mode-fluct}, appears. The structure of this term
reflects the fact that due to the presence of one uncompensated site
the global $\GL$-invariance of the model is lost. ($S_{{\rm
    top},2}[T]$ is not invariant under $T\to T_L T T_R$ even for
constant $T_{L,R}$.) In the majority of cases this extra term is of
little interest. However for the spectral properties of a short
isolated system, the presence of $S_{{\rm top,2}}$ bears crucial
effects. Indeed we will see that this term is responsible for the
formation of staggering phenomena akin, and probably related to the
effects discussed earlier in section \ref{sec:cond-short-syst-1}.

\begin{figure}[htbp]
  \begin{center}
    \epsfig{file=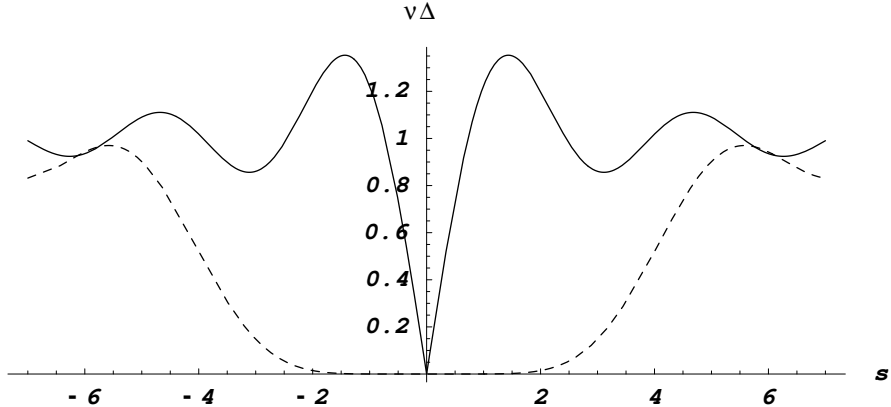,height=5.5cm} \\[0.5cm]
    \caption{Density of states of the short sublattice system. Solid
      line: DoS of a system with $L$ even. Dashed line: DoS of a
      system with $L$ odd and $N=3$. (The singular zero energy spike
      causing the spectral depletion up to values $s \sim 3$ is not shown.)}
    \label{fig:figure2}
  \end{center}
\end{figure}

Adding the two contributions to the action and using Eq. (\ref{eq:11})
we obtain
$$
\nu(\epsilon) = {1\over 2 \Delta} \int dT\; (T_{11}+T^{-1}_{11})
e^{i{s\over 2} (T+T^{-1}) - N \strln(T)},
$$
for the $\sigma$-model representation of the zero-dimensional DoS.
The task thus is to integrate over a single copy of the target
manifold $\GL$. For sufficiently small energies $\epsilon \sim
\Delta$, the action is not large enough to confine the integrand to
the origin of the group manifold implying that the integral has to be
done non-perturbatively.  Referring to Ref. \cite{altland99:NPB_flux}
and Appendix \ref{sec:cond-short-syst} for technical details we here
merely display the final result of this integration procedure,
\begin{eqnarray}
  \label{eq:36}
 \nu(s) = {\pi s \over 2\Delta}
\left[ J_0^2(s) + J_1^2(s)\right],\qquad &&L\;\mbox{even}\nonumber\\
  \nu(s) = N\delta(\epsilon) + {\pi s\over 2\Delta}
\left[ J_N^2(s) -
J_{N+1}(s)J_{N-1}(s)\right],\qquad&& L\;\mbox{odd}
\end{eqnarray}
The structure of these DoS profiles is shown in Fig.~\ref{fig:figure2}
for the example $N=3$.

Eqs. (\ref{eq:36}) have been obtained earlier within pure random
matrix theory\cite{verbaarschot93}, and its supersymmetry
implementation\cite{ast,verbaarschot96:susy}. Studies of chiral random
matrix ensembles were largely motivated by the relevance of the
spectral structure of effectively zero dimensional chiral systems in
{\it finite size} lattice QCD. More generally, 'microgaps' of the type
shown in Fig.~\ref{fig:figure1} are an omnipresent side effect seen in
the spectrum of generic chiral systems with finite mean level spacing
$\Delta$.

Qualitatively, origin and structure of these gaps can easily be
understood. First, the chiral symmetry $[H,\sigma_3]_+=0$ entails that
for any {\it non-vanishing} energy level $\epsilon_n$ its negative
$-\epsilon_n$ is an eigenvalue, too. Disorder generated level
repulsion prevents these states from coming close to each other (on
the scale of the mean level spacing) which explains the presence of
the spectral gap in the $L$-even case.  For $L$ odd, this picture has
to be modified. To understand what is happening, let us imagine the
Hamiltonian as a block off-diagonal matrix in sublattice space:
$$
H = \left(\matrix{&Z\cr Z^\dagger&}\right)\qquad
\begin{array}{l}
\mbox{odd}\cr\mbox{even.}
\end{array}
$$
Since $L$ is odd, the number of odd sites exceeds the number of
even sides by one, i.e. the blocks $Z$ are rectangular with $N
(L-1)/2$ rows and $N(L+1)/2$ columns. Now, any block-off diagonal
matrix with $k$ rows and $l$ columns has $|k-l|$ eigenvalues $0$.
Applied to our system, this means that the $L$ odd system has $N$ zero
modes for any realization of disorder (the $\delta$-function
contribution in (\ref{eq:36}).)  Other levels repel from this
concentrated accumulation of spectral weight which explains the
bathtub type suppression of the DoS up to energies $\sim N\Delta$.

Summarizing we have seen that the $L$ even/odd staggering behaviour
observed earlier in  connection with the conductance pertains to
spectral properties. Above we had argued that the vanishing of the
conductance in the $L$ even case was due to the peculiar spatial
profile of zero energy wave functions. In view of (\ref{eq:36}) it is
tempting to relate the same effect to the vanishing of spectral weight
at zero energy, although we have not analysed this picture any
further. 

Keeping in mind the tendency towards buildup of zero energy spectral
weight in systems with mis-matched sublattice structure we next
turn back to the analysis of large systems.

\subsection{Heat kernel for finite energies}
\label{sec:heat-kernel-finite}

The aim of this section is to compute the DoS of a large system with
$L\gg \xi$. To facilitate comparison with the behaviour of the
conductance we will drop the assumption of isolatedness and again
couple the system to leads. Notice that this stands complementary to
the analysis of Ref. \cite{brouwer00_off}, where a closed system was
discussed. That we will obtain identical scaling of the DoS is proof
of the (in view of the existence of topological zero modes not
entirely obvious) assertion \cite{brouwer00_off} that boundary
conditions do not affect the bulk spectrum.

We will not be able to compute the DoS $\nu(\epsilon)$ for arbitrary
$\epsilon$.  Instead, an asymptotic expression valid for low energies
will be derived. Remembering the behaviour found for small systems 
we anticipate nonanalytic behaviour, $\nu(\epsilon) \sim
\ln^n(|\epsilon|) |\epsilon|$, where the logarithmic factor is due to
potentially existent localization corrections to the zero mode
behaviour. Our objective is to identify the most singular
contribution of this type.

Starting point of the analysis is the transfer matrix representation
(\ref{eq:13}) of $\nu(\epsilon)$. To evaluate this expression we need
to compute the functions $Y^{L/R}$, now for finite potential $V$. We
will do this following a procedure developed in Ref.~\cite{AltF}.
Starting out from a spectral decomposition of the type (\ref{eq:39}),
we first notice that only zero energy eigenfunctions contribute to
long distance behaviour of $Y_{L/R}(r)$.  This means that our 'time'
dependent 'Schr\"odinger equation' (\ref{eq:17}) can effectively be
replaced by the stationary form
\begin{equation}
  \label{eq:40}
t\gg \xi:\qquad \left(-(\D\pm\A)^2 + \eta (\cosh(x)-\cos(y))
\right) Y_{L/R}(x,y)=0,  
\end{equation}
where we have substituted the explicit form of the potential and
omitted the spatial argument of $Y_{L/R}$ for simplicity. For later
convenience we have also analytically continued from real to imaginary
energy arguments, $-is^+ \to 4 \eta >0$.  Substitution of these
functions into Eq. (\ref{eq:13}) yields the reduced representation
\begin{eqnarray}
\label{eq:41}
\nu(E) = \nu_0\left(1+ {\,\rm Re\,}  {1\over 2\pi}
\int dxdy\;{\cosh(x)-\cos(y)\over \sinh^2 \left({x-iy\over
      2}\right)}
Y_L(x,y)Y_R(x,y)\right).
\end{eqnarray}
To obtain this equation we have expanded the pre-exponential sources
in Grassmann variables and integrated over these. The constant
contribution $\nu_0$ appears as a consequence of the Efetov-Wegner
theorem. (The representation of the pre-exponential term in polar
coordinates contains a purely non-Grassmann contribution. Integration
over a term of this type obtains the integrand at the
origin\cite{Efetbook} which, in our case, equals unity.)

We now turn to the actual computation of the functions $Y_{L/R}$.  The
first step is the derivation of a set of matching, or boundary
conditions relating the heat kernel to its $\eta\to0$ asymptotics.
To this end we evaluate the $\eta=0$ spectral decomposition
(\ref{eq:39}) in the limit $t\gg \xi$. Neglecting contributions that
decay exponentially in $t/\xi$, it is straightforward to obtain the
asymptotic expressions:
\begin{eqnarray*}
&&  Y_{L \atop R}(x,y) \stackrel{\eta\to 0}{\longrightarrow} 1,
  \qquad N\;\mbox{even},\nonumber\\
&&  Y_{L \atop R}(x,y) \stackrel{\eta \to
    0}{\longrightarrow}{1\over 2} (e^{\pm(x-iy)}+1),
  \qquad N\;\mbox{odd},
\end{eqnarray*}
where we have temporarily neglected the staggering parameter $f$. The
structure of these functions will be motivated shortly.  Turning to
the case $\eta\not=0$, we next gauge and transform Eq.~(\ref{eq:40}) as in section \ref{sec:spectr-eigenf-kinet}.
Transformation $ Y_{L\atop R}(x,y) \to
J^{1/2}(x,y)e^{\pm{1-{\cal P}_N\over 2}(x-iy)}Y_{L\atop R}(x,y)\equiv
\tilde Y_{L \atop R}(x,y)$, where ${\cal P}_N \equiv (1+(-)^N)/2$
brings the equation into the form 
\begin{equation}
  \label{eq:37}
  \left(\partial^2_x+\partial^2_y - \eta (\cosh(x)
    -\cos(y))\right) 
\tilde Y_{L/R}(x,y)=0,
\end{equation}
while the transformed $\eta\to 0$ asymptotics read
\begin{eqnarray}
\label{eq:38}
  \tilde Y_{L/R}(x,y) \stackrel{\eta\to 0}{\longrightarrow}&
\sinh^{-1}\left({x-iy\over 2}\right) &
\stackrel{|x|\gg 1}{\longrightarrow} 2\sgn(x) e^{-\sgn x
  \left({x-iy\over 2}\right)},
  \qquad N\;\mbox{even},\nonumber\\
  \tilde Y_{L/R}(x,y) \stackrel{\eta \to
    0}{\longrightarrow}&\coth\left({x-iy\over 2}\right)&\stackrel{|x|\gg
    1}{\longrightarrow}\sgn(x),
  \qquad N\;\mbox{odd}.
\end{eqnarray}
Here we have anticipated that the dominant contribution to the above
double integral representation of the DoS will come from large values
of the non-compact variable $x$. Eqs. (\ref{eq:37}) and (\ref{eq:38})
have the nice feature of full separability. Writing $\tilde Y(x,y) =
{\cal N} Y_1(x) Y_2(y)$, where ${\cal N}$ is a normalization factor
and the subscript $L/R$ has been dropped for simplicity, Eq.
(\ref{eq:37}) can be traded for the set of decoupled ordinary
differential equations
\begin{eqnarray*}
&& \left( \partial_x^2 - {\eta\over 2} e^{x}- {{\cal P}_N\over 4}\right)Y_1(x)
= 0+ {\cal O}(\eta),\\
&&  \left(\partial_y^2 + {{\cal P}_N\over 4}\right) Y_2(y) = 0 + {\cal
  O}(\eta).
\end{eqnarray*}
The second equation is trivially solved by $Y_2(y) \approx e^{\pm
  i{\cal P}_N{y\over 2}}$. The first line is a Bessel equation. Its
two solutions are given by $I_{{\cal P}_N}((2\eta)^{1/2}e^{|x|/2})$
and $K_{{\cal P}_N}((2\eta)^{1/2}e^{|x|/2})$. Discarding the
exponentially divergent solutions $I_\nu$ and using that for small
arguments, $K_0(z) \approx - \ln(z)$ and $K_1(z) \approx z^{-1} +
{z\over 2}\ln\left( {z\over 2}\right)$, the normalization constants
${\cal N}$ can now be fixed by matching to the zero energy
asymptotics in the limit of small $\eta$. This obtains the approximate
solution
\begin{eqnarray*}
&  \tilde Y_{L/R}(x,y) \stackrel{|x|\gg 1}\approx (8\eta)^{1/2} K_1
\left((2\eta)^{1/2}e^{|x|/2}\right)e^{i{\,\sgn\,}x
  {y\over 2}}, &\qquad N {\;\rm even},\\
&\tilde Y_{L/R}(x,y) \stackrel{|x|\gg 1}\approx -{2\over \ln \eta}
K_0\left((2\eta)^{1/2}e^{|x|/2}\right),&\qquad N {\;\rm odd}.
\end{eqnarray*}
It is now a straightforward matter to substitute this expression back
into the above integral representation for $\nu$ and to integrate over
coordinates. The $y$-integration, extending over a purely harmonic
integrand, is trivially done. (Notice that the $\sinh^{-2}$-factor in
(\ref{eq:41}) cancels against the factor $(J^{1/2})^2$ from the
similarity transformation). As for the $x$-integration, we note that due
to the exponentially decaying asymptotics $K_\nu(z) \sim
\exp(-z/2)$ for $|z|\gg 1$, the integral can be cut off at
$(2\eta)^{1/2}e^{|x|/2} \sim 1 \Rightarrow |x|\sim -\ln(2\eta)$. Within
the domain of integration, the Bessel functions can be replaced by the
small argument asymptotics specified above. Substituting these
expressions it is then straightforward to obtain $\nu(\eta) \approx \nu_0
{\,\rm Re\,}\eta \ln^2\eta$ (even $N$) and $\nu(\eta)\approx \nu_0
{\,\rm Re\,} (\eta \ln^2 \eta)^{-1}$ (odd $N$) for the small
$\eta$-asymptotics of the DoS. Analytic continuation back to real
energies finally leads to the result
\begin{eqnarray}
  \label{eq:43}
  \nu(s) \approx - \nu_0 |s| \ln|s|,\qquad N \,\mbox{even},&&\nonumber\\
  \nu(s) \approx - {\nu_0\over |s|\ln^3|s|},\qquad N
    \,\mbox{odd},&&
\end{eqnarray}
for the low energy behaviour of the DoS.  Eqs. (\ref{eq:43}) agree
with the results found earlier in Ref. \cite{brouwer00_off}.  

We finally discuss the extension of the above results to non-zero
staggering. For non-vanishing $f$ and even $N$, the large argument
asymptotics (\ref{eq:38}) generalize to
\begin{equation}
  \label{eq:45}
  \tilde Y_{L/R}(x,y) \stackrel{\eta\to 0,x\gg 1}{\longrightarrow}
2 e^{{1\over 2}
  (\pm f +1)(x-iy)}.  
\end{equation}
One can now follow the same steps as in the non-staggered cases above
to  obtain the result
\begin{equation}
  \label{eq:44}
\nu=2\nu_0 {\Gamma(|f|)\over \Gamma(2+|f|)}
{2^{-2(1-|f|)}\over|f|}
\cos\left({\pi\over 2} (1-|f|)\right) |s|^{1-|f|}.  
\end{equation}
For non-zero $f$, the DoS vanishes in a more singular manner as in the
non-staggered case. This behaviour provokes the question, how matching
with the diverging profile in the case $N$ odd, $f=0$ might be
obtained. The principal structure of the theory entails that ($N$
even/$f=1$) should be equivalent to ($N$ odd/$f=0$). On the other
hand, the $f\nearrow 1$ version of Eq. (\ref{eq:44}) certainly does
not agree with the divergent result (\ref{eq:43}).

To resolve this paradox, it is helpful to re-interpret the asymptotic
expressions (\ref{eq:38}) and (\ref{eq:45}) within the
quantum mechanical picture of the theory. Focusing on the compact
sector and temporarily ignoring the factor $J^{1/2}$ from the
transformation to a flat Laplacian, these functions acquire the
meaning of ground state wave functions $\Psi_0$ of a one dimensional
ring subject to a gauge flux ${N+f\over 2}$. For $N$ even and zero
$f$, an integer number of 'flux quanta' pierce the ring, and the
ground state wave function carries zero 'momentum', $\Psi_0(y) \propto
1$ which, after multiplication with $J^{1/2}$ leads to the first line
of Eq.  (\ref{eq:38}). In contrast, for $N$ odd and $f$ still zero, a
half integer flux pierces the ring. This is a special situation in the
sense that the ground state wave function is two-fold degenerate, i.e.
$\Psi_0(y)= c_+ e^{iy/2} + c_- e^{-iy/2}$. Our
earlier analysis has fixed the a priori un-determined constants $c_\pm$
to a symmetric configuration.  Inclusion of the non-compact variable
and multiplication with $J^{1/2}$ then leads to the second line of
(\ref{eq:38}).

We can now understand what happens as $f$ is turned on for an $N$ even
configuration: a flux $f/2$ is sent through the ring and the ground
state wave function remains unique (cf. Eq. (\ref{eq:45})) {\it until}
$f$ comes close to the critical value $1$. In the immediate vicinity
of the degeneracy point, the fact must no longer be neglected that our
one-dimensional system is subject to a weak potential $\eta \cos(y)$.
For values of $f$ such that the level splitting $\sim 1-f$ between the
two nearly degenerate levels becomes comparable with the
characteristic strength of the potential $\sim \eta$, the 'true'
ground state configuration is given by the symmetric superposition of
the two levels, as in the $N$ odd case.  For these values of $f$ the
ground state configuration is given by the second line of Eq.
(\ref{eq:38}) and the DoS follows the $N$ odd asymptotics.

This qualitative argument predicts that for asymptotically small
energies, the DoS scales as in Eq.  (\ref{eq:44}). However, for larger
values of the energy, $\eta \sim |1-f|$, a crossover to the
characteristics of the $N$ odd $f=0$ DoS profile takes place.

\section{Summary}
\label{sec:summary}

In this paper, transport and spectral properties of weakly disordered
quantum sublattice wires have been explored from a fieldtheoretical
perspective. We re-derived results obtained previously within the DMPK
transfer matrix formalism, observed a surprisingly strong sensitivity
of system properties to the realization of the lead/device coupling,
and found that conductance and DoS, at least of short systems, exhibit
drastic dependence on the parity of the total site number in the
sublattice chain. It is likely that both this phenomenon and the
dependence of system properties on the parity of the channel number
root in the same origin, i.e. the existence of zero energy states for
effectively block off-diagonal Hamiltonians with rectangular
(non-quadratic) block structure. Although we are not aware of an
intuitive explanation for the channel number parity effects, this
belief is supported by the observation that all staggering phenomena
are controlled by the same topological term $S_{\rm top}$ in the
action of the $\sigma$-model. From these findings one might expect
that the delocalized band-centre behaviour exhibited by the {\it
  two}-dimensional sublattice model, is driven by the $2d$-analogue of
this operator.  Curiously, this is not so. In the two-dimensional
field theory, the so-called Gade term $S_{\rm Gade}$, a two-gradient
operator contribution with small and non-universal coupling constant,
drives the system towards de-localization. The operator $S_{\rm top}$
does have a generalization to two dimensions\cite{altland99:NPB_flux},
but its role has not been investigated so far. Summarizing we find
that comparable phenomenology (metallic behaviour and diverging DoS)
in one and two dimensions is described by different operators
in the $\sigma$-model action.  This indicates that some 'deeper'
physical principle, not understood at present, lies beyond the visible
structure of the field theory.

{\it Acknowledgement:} Discussions with P. Brouwer, J. Chalker, V.
Kravtsov, A. Ludwig, C. Mudry, B.  Simons, and J. Verbaarschot are
greatfully acknowledged.  Special thanks to M.  Zirnbauer for
discussions, numerous supportive hints, and for drawing our attention
to the most important aspect of the field theory, the topological
term.  This work was partly supported by Sonderforschungbereich 237 of
the Deutsche Forschungsgemeinschaft.

\appendix

\section{Field Integral Formulation}
\label{sec:field-integr-form}

$\sigma$-model representations of zero- and two-dimensional systems
with $A$III-symmetry have been constructed in different contexts
before\cite{gade93,verbaarschot96:susy,altland99:NPB_flux,guruswamy00}.
That the following two sections discuss the construction of the field
theory in some detail is motivated by non-generic features particular
to the $1d$-system, most notably the coupling operators and the
existence of topological structures.  In the present section, we will
derive a representation of the model and its correlations functions in
terms of a supersymmetric field integral.  The projection of this,
a priori exact representation onto its low energy sector will be the
subject of the subsequent Appendix \ref{sec:deriv-field-theory}.

Consider the two correlation functions
\begin{eqnarray*}
 & C^{(1)}_{\alpha \alpha'}\equiv \langle G_{\alpha\alpha'}(E^+) \rangle
 &\qquad \mbox{DoS}\\ 
 & C^{(2)}_{\alpha \beta \alpha' \beta'}\equiv \langle 
G_{\alpha \beta}(0^+) G_{\alpha' \beta'}(0^-)
 \rangle&\qquad\mbox{conductance}, 
\end{eqnarray*}
relevant for the computation of DoS and conductance, respectively.  To
compute these objects, we follow the now standard supersymmetry scheme
for disordered electronic systems\cite{Efetbook} and represent the
Green functions as
$$
G_{\alpha \alpha'}(z) = \left. {\delta \over \delta J^{\rm b}_{\alpha' \alpha}}\right|_{\hat
  J=0}Z[\hat J] = - \left.
{\delta \over \delta J^{\rm f}_{\alpha '\alpha}}\right|_{\hat
  J=0} Z[\hat J],
$$
where 
\begin{equation}
  \label{eq:3}
Z[\hat J] = \int {\cal D}(\bar\psi,\psi) 
e^{i  \bar \psi \left(z-\hat H + i\pi \sum_C
  W^{C}W^{CT}- \hat J\right)\psi},
\end{equation}
and we have assumed that ${\rm sgn\,Im\,} z>0$. (In the opposite case,
the sign of both the total action and the coupling operator change.)
Here $\psi = \{(S_\alpha,\chi_\alpha)^T\}$ and $\bar \psi = \{(\bar
S_\alpha,\bar\chi_\alpha)\}$ are two-component superfields where $\bar
S_\alpha$ is the complex conjugate of $S_\alpha$, while $\chi_\alpha $
and $\bar \chi_\alpha$ are independent Grassmann variables. The source
field
$$
\hat J = \left(\matrix{\hat J^{\rm b}&0\cr 0&\hat J^{\rm f}}\right)
$$
where $\hat J^{\rm b,f} = \{\hat J^{\rm b,f}_{\alpha \alpha'}\}$ are
ordinary matrices in site and orbital space.

From this representation, the one-point correlation function obtains
as $C^{(1)}_{\alpha \alpha'}= \left. \delta_{J^{\rm b}_{\alpha'
      \alpha}}\right|_{\hat J=0}\langle Z[\hat J]\rangle$ for $z=E^+$.
However, at first sight it looks like {\it two} Gaussian field
integrals (\ref{eq:3}) are needed to compute the two-particle
correlator $C^{(2)}$, one for the Green function $G(0^+)$ the other
for $G(0^-)$. Fortunately, this is not so, a direct consequence of
the chiral symmetry of the Hamiltonian: the relation $[\sigma_3,\hat
H]_+=0$ implies that
$$
\hat G(z) = - \sigma_3 G(-z) \sigma_3
$$
and thus $G_{\alpha \beta}(0^+) = (-)^{l+k+1} G_{\alpha
  \beta}(0^-)$, where $l$ and $k$ are the site indices carried by the
composite variables $\alpha$ and $\beta$, respectively.  In other
words, the retarded and the advanced Green function are not
independent and we can obtain the correlation function $C^{(2)}$ as
$$
C^{(2)}_{\alpha \beta \alpha' \beta'} = (-)^{l'+k'}
\left. {\delta^2
    \over \delta J^{\rm b}_{\beta \alpha} J^{\rm
      f}_{\beta'\alpha'}}\right|_{\hat 
  J=0}\langle Z[\hat J]\rangle
$$
from the comparatively simple generating functional
of the one-point function (evaluated
for $z=0^+)$.

\section{Derivation of the field theory} 
\label{sec:deriv-field-theory}

In this Appendix we derive the effective Lagrangian (\ref{eq:9}) from
the basic representation (\ref{eq:3}). To keep the discussion
simple, we will suppress the source-field dependence of the partition
function in much of what follows. The final results for the
correlation functions $C^{(1,2)}$, obtained by straightforward
expansion of the action to first and second order in $\hat J$ are
displayed in the final Eq. (\ref{eq:10}).

The derivation of the effective action essentially follows the
standard \cite{Efetbook} construction route of field theories of
disordered electronic systems. We begin by averaging the partition
function over the Gaussian distribution of the random hopping matrices
$R$:
$$
Z[0] = \int {\cal D}(\bar\psi,\psi) 
e^{i  \bar \psi \left(z-\hat H_0 + i\pi \sum_C
  W^{C}W^{CT}\right)\psi -  {\lambda^2 \over N} \sum_l {\rm \,str\,}(\psi_{l\mu} 
\bar \psi_{l\mu} \psi_{(l+1)\nu}\bar \psi_{(l+1)\nu})},
$$
where $\hat H_0 = \{t_{ll'} \delta_{\alpha \alpha'}\}$ is the clean
part of the Hamiltonian. Next, the quartic contribution is decoupled by
means of two auxiliary fields:
\begin{eqnarray*}
&&Z[0] = \int {\cal D}(Q,P)
e^{-N\sum_l {\,\rm str\,}(Q_{l,l+1}^2 + P_{l,l+1}^2)}
\int {\cal D}(\bar\psi,\psi)
e^{i   \bar \psi \left(z-\hat H_0 + i\pi \sum_C
  W^{C}W^{CT}\right)\psi}\times\\
&&\hspace{2.0cm} \times e^{ 
i\lambda \sum_{l,{\rm even}} \bar\psi_l
\left[Q^+_{l,l+1}+Q^+_{l,l-1}\right]\psi_l  
-i\lambda \sum_{l,{\rm odd}} \bar\psi_l
\left[Q^-_{l,l+1}+Q^-_{l,l-1}\right]\psi_l  
}, 
\end{eqnarray*}
where $Q^\pm = Q \pm iP$, and $Q$ and $P$ are two-component
supermatrix fields (reflecting the two-component matrix structure of
the dyadic products $\psi \bar \psi$) living on the non-directed {\it
  links} of our system. Both, internal structure and symmetry
properties of these fields will be discussed momentarily. At this
stage, we merely anticipate that the field configurations relevant to
the long range behaviour of the system will be smooth. The structure
of the action then suggests to define $Q^\pm_l \equiv {1\over
  2}(Q^\pm_{l,l+1} + Q^\pm_{l,l-1})$ as a new field variable, which
now again lives on the sites of our system.

To account for the staggered even/odd structure of the theory we next
'double the unit cell' and define a two component field
$$
\Psi_j = \left(\matrix{\psi_{2j+1}\cr \psi_{2j}}\right)\equiv
\left(\matrix{\Psi_{1,j}\cr \Psi_{2,j}}\right)
$$
Here we have introduced a new counting index $j=0,\dots,L/2$
enumerating the doubled unit cells of our system. To avoid confusion,
we will systematically designate the index $0,\dots,L$ of the
'primitive' sites by $l,l',\dots$ and the new index by $j,j',\dots$.
Expressed in terms of $\Psi$ the functional integral assumes the form
\begin{eqnarray*}
&&Z[0] = \int {\cal D}(Q,P)
e^{-2 N\sum_j {\,\rm str\,}(Q_j^2 + P_j^2)}\int {\cal
  D}(\bar\Psi,\Psi)\times\\ 
&&
\times
\exp\bigg( i  \bar \Psi \bigg(z + 2
\lambda(Q+iP\sigma_3)+
 i\pi \sum\limits_{C=L,R}
  W^{C}W^{CT} +
\sigma_+ H^{12}_{0}
+\sigma_- H^{21}_{0}\bigg)\Psi\bigg).
\end{eqnarray*}
Here, $\sigma_\pm = \sigma_1 \pm \sigma_2$ where
$\sigma_i, i=1,2,3$ are Pauli matrices acting in the space defined
through the two-component structure of $\Psi$. In lack of better
terminology, we will refer to this space as the 'chiral space'.  The
second line of the expression above, purely off diagonal in the chiral
space, contains the clean Hamiltonian. The explicit lattice structure
of the blocks $H_0^{21}=(H_0^{12})^\dagger$ is given by
\begin{eqnarray*}
&&H_{0,ll'}^{12}= \delta_{ll'}(1+a) + \delta_{ll'-1}(1-a),\\
&&H_{0,ll'}^{21}= \delta_{ll'}(1+a) + \delta_{ll'+1}(1-a),
\end{eqnarray*}
where $a$ is the staggering parameter introduced in
(\ref{eq:1}). Finally, notice that the chiral matrix structure of the
coupling operator is given by
$$
W^C_j W^{CT}_j = \left(\matrix{W^C_{2j+1} W^{CT}_{2j+1}&\cr
&W^C_{2j} W^{CT}_{2j}}\right).
$$
At this stage, the superfield $\Psi$ can be integrated out and we
arrive at
\begin{eqnarray*}
&&Z[0] = \int {\cal D}(Q,P)
e^{-2 N\sum_j {\,\rm str\,}(Q_j^2 + P_j^2)}\times\\
&&\hspace{2.0cm}
\times
\exp \left(-N {\rm \, str\,\ln}\bigg(z+ i\pi \sum\limits_C
  W^{C}W^{CT} + 2\lambda (Q+iP\sigma_3)+
\sigma_+ H^{12}_{0}
+\sigma_- H^{21}_{0}\bigg)\right).
\end{eqnarray*}
To make further progress, we subject the functional integral to a
saddle point analysis.  Following the standard scheme\cite{Efetbook},
we seek for saddle point configurations $\bar Q$ and $\bar P$ that are
matrix-diagonal and spatially uniform. Further, we temporarily set
$a=W={\, \rm Re\,}z=0$ and neglect boundary effects due to the finite
extent of the system. Making the ansatz $\bar Q = i q \cdot \openone$,
$\bar P = p\cdot \openone$, where $q$ and $p$ are complex numbers and
$\openone$ is the two-dimensional unit matrix in superspace, a
variation of the action w.r.t. $\bar Q$ and $\bar P$ then generates
the set of equations
\begin{eqnarray*}
q &=& {i\lambda\over 2} {\rm tr\,}\left(\hat G_0^{-1} + 2 i \lambda(q+p
  \sigma_3)\right)_{jj}^{-1}\\ 
p &=&  {i\lambda\over 2} {\rm tr\,}\left[\left(\hat G_0^{-1} +
    2i \lambda(q+p\sigma_3 )\right)_{jj}^{-1}\sigma_3\right], 
\end{eqnarray*}
where $G_0 \equiv \left. (i\delta + \sigma_+ H^{12}_{0} +\sigma_-
  H^{21}_{0})^{-1}\right|_{a=0}$ and the trace extends over the two
chiral components of the operators (but not over superspace). To
evaluate the trace, we Fourier transform to momentum space.  Defining
the transform through
\begin{eqnarray*}
&&f(k) = \left({2\over L}\right)^{1/2} \sum_j e^{i2kj}f_j\\    
&&f_j =  \left({2\over L}\right)^{1/2} \sum_k e^{-i2kj}f(k)    
\end{eqnarray*}
(where the factor of two in the exponent serves as a mnemonic
indicating that we have doubled the unit cell of our system) one finds
that the Green function $G_0$ is diagonal in momentum space with $
G_0(k) = \left(i\delta + (1 + e^{2ik}) \sigma_+ +
  (1+e^{-2ik})\sigma_-\right)^{-1} $. It is now straightforward to
verify that our equations are solved by $p=0$ and $q$ determined
through the self consistency equation
$$
q= 2\lambda^2 {2\over L}\sum_k {1\over (2\lambda q)^2 + 2(1-\cos(2k))}.
$$
Replacing the sum by an integral, $\sum_k \to {L\over 2\pi}
\int_0^\pi dk$, one finds that the equation is solved by $q=\lambda
/2$. 

The presence of the chiral symmetry implies that the configuration
$\bar Q = {i\lambda \over 2}\openone$ is but a particular
representative of a whole manifold of solutions. To explore the
morphology of that manifold, we notice that our restricted ($z=W=0$)
action is invariant under the transformation
\begin{eqnarray*}
\bar \Psi_1 \to \bar \Psi_1 T_1^{-1},&\qquad&
\bar \Psi_2 \to \bar \Psi_2 T_2,\\   
\Psi_1 \to \bar T_2^{-1} \Psi_1,&\qquad&   
\Psi_2 \to  T_1\Psi_2,\\
Q+iP \to T_1(Q+iP)T_2,&\qquad&Q-iP \to T_2^{-1}(Q-iP)T_1^{-1},
\end{eqnarray*}
where $T_1,T_2 \in \GL$. This symmetry, the super-generalization of
the fermion symmetry (\ref{eq:42}), states that the model has $\GL
\times \GL$ as a global invariance group. Applying the transformation
to the diagonal stationary phase solution discussed above we find that
$$
{i \lambda \over 2} \openone \to {i\lambda \over 2}
\left(\matrix{T&\cr&T^{-1}}\right), 
$$
where $T=T_1T_2$. Arguing in reverse, we conclude that any matrix
$T$ defines a solution of the mean field equations, i.e. ${\rm
  GL}(1|1)$ is the Goldstone mode manifold of the model. This is the
celebrated mechanism of chiral symmetry breaking (see
Ref. \cite{verbaarschot00} for review): field theory implementations of
models with a discrete chiral symmetry on the microscopic level,
possess continuous factor groups ${\rm G}\times {\rm G}$ as
symmetry manifolds. (In our case ${\rm G}={\rm GL}(1|1)$.)  This
symmetry is spontaneously broken by the saddle point configurations of
the model. What remains is a Goldstone mode isomorphic to a single
factor ${\rm G}$.

Combining these results, we parameterize our field manifold as
$$
\left(\matrix{Q+iP&\cr&Q-iP}\right) = {i\lambda \over 2}
\left(\matrix{UT&\cr&T^{-1}U}\right),
$$
where both $T,U \in {\rm GL}(1|1)$. Here, the matrices $T$ span the
Goldstone model manifold whereas the $U$'s, incompatible with the
global symmetry of the model, represent massive modes.  The next
logical step in the construction of the field theory is to substitute
these configurations back into the action and to expand in (i) energy
arguments $z$ and matrix elements $W$, (ii) long-ranged spatial
fluctuations of the Goldstone modes, and (iii) massive modes.

\subsection{Goldstone mode fluctuations}
\label{sec:goldst-mode-fluct}

We begin with the second element of the program formulated above, the
expansion of the action in long ranged spatial fluctuations of the
Goldstone mode.  Temporarily setting $z=W=0$, $U=\openone$, we
re-organize the 'str ln' of the action according to
$$
X\equiv N{\rm str\,\ln}\left(\matrix{i\lambda^2 T &
    H_0^{12}\cr H_0^{21} & i\lambda^2 T^{-1}}\right) = N{\rm
  str\,\ln}\left(\matrix{i\lambda^2 & H_0^{12}\cr T
    H_0^{21}T^{-1} & i\lambda^2}\right),
$$
where $T$ is a slowly fluctuating field of Goldstone modes. Writing
$T H_0^{21}T^{-1} = H_0^{21} + T [H_0^{21},T^{-1}]$ and noticing that,
due to the slow fluctuation of $T$, the commutator is small, we expand
as
$$
X= N {\rm str\,}(\bar G^{12}T [H_0^{21},T^{-1}]) - {N\over 2} {\rm
  str\,}(\bar G^{12}T [H_0^{21},T^{-1}]\bar G^{12}T
[H_0^{21},T^{-1}])+ \dots,
$$
where we have defined $\bar G_0 = (G_0 + i\lambda)^{-1}$ and the
ellipses stand for infrared irrelevant higher order commutator terms.
One next Fourier transforms these expressions, substitutes the
explicit momentum representation $G_0(k)$ and uses that the
characteristic momentum $q$ carried by the transforms $T(q)$ is small.
The subsequent integral over the 'fast momentum' $k$ is most
economically done by noticing that full integration over $k$ amounts
to integrating the characteristic phases $\exp(i2k)$ once over the
complex unit circle. This integral has a simple pole inside the
integration contour, whose residues depend on the 'small' momentum
$q$. Expanding the residues to lowest non-vanishing order in $q$ and
transforming back to coordinate space one obtains $X= S_{\rm top} +
S_{\rm fl}$, where the two contributions are displayed in Eq.
(\ref{eq:4}) and a continuum limit $\sum_j \to {1\over 2}\int_0^L dr$
has been taken.

Notice that the above discussion implicitly assumed that the number of
sites of our system is even: the operator $X$ had a structure where
each $T$ (living at an odd site) came with a partner $T^{-1}$ at the
neighbouring even site. For a system with an {\it odd} number of sites
$L$, however, there remains one uncompensated degree of freedom
$T((L-1)/2)$ at the terminating site. Neglecting gradients, the action
due to this extra contribution reads $S_{{\rm top},2} = N \strln
T(L)$. In principle, this contribution is as relevant as the 'bulk'
contribution to the topological action: it is local in space but, on
the other hand, does not contain derivatives. Indeed, the structure of
$S_{{\rm top},2}$ is closely related to that of $S_{\rm top}$ as can
be seen by representing the latter as a boundary action (cf. the third
line of (\ref{eq:35}).) Yet in the majority of cases, the extra
contribution $S_{{\rm top},2}$ does not play much of a role, wherefore
we have largely ignored it in main analysis. E.g., for a system coupled
to the outside world, $S_{{\rm top},2}=N\strln(T(L))=N2\pi i$,
evaluates to a phase of no physical effect (cf. the related discussion
around Eq. (\ref{eq:35}).)  There is, however, one exception to that
rule, viz. the physics of {\it isolated} short systems with an odd
number of sites discussed in section \ref{sec:density-states-short},
where the presence of the extra contribution is of key relevance.

\subsection{Finite energies and coupling to the leads}
\label{sec:finite-energ-coupl}

To obtain the action associated to finite $z$ and $WW^T$, we organize the
action as,
\begin{eqnarray*}
&& {\,\rm str\,\ln}\left(
\matrix{z+i\pi WW^T + i \lambda^2 T & H_0^{12}\cr
H_0^{21}&z+i\pi WW^T + i\lambda^2 T^{-1}}\right) \approx \\
&&\hspace{2.0cm}\approx  {\,\rm str\,\ln}\Bigg[\left(
\matrix{(z+i\pi WW^T) T^{-1}&\cr&(z+i\pi WW^T)T&}\right)
+ \underbrace{\left(\matrix{i \lambda^2  & H_0^{12}\cr
H_0^{21}& i\lambda^2}\right)}_{\displaystyle \bar G^{-1}}\Bigg]\approx\\.
&&\hspace{2.0cm}\approx  \sum_j {\,\rm str\,}
\left[(z+i\pi (WW^T)_{2j+1}) T_{l}^{-1}\bar G^{11}_{jj} + (z+i\pi
  (WW^T)_{2j})T_l\bar G^{22}_{jj}\right]=\\
&&\hspace{2.0cm}= -i  z \pi\nu_0\sum_j {\,\rm
  str\,}(T+T^{-1}) + {\pi \over 2}  \sum_j {\,\rm \,str\,}
\left[(WW^T)_{2j+1} T_{l}^{-1}+ 
  (WW^T)_{2j}T_l\right],
\end{eqnarray*}
where we have used that (from the saddle point equation) $\nu =
i{N\over \pi} \bar G^{11}_{jj}=i {N\over \pi} \bar G^{22}_{jj} =
{N \over \pi}$ and $\nu$ is the DoS per site evaluated for energies far
away from the middle of the band.  We next assume that in the coupling
region to the leads, i.e.  the region where the envelope function $f$
defined through (\ref{eq:6}) is non-vanishing, fluctuations of the
Goldstone modes are negligible.  Application of the orthogonality
relation (\ref{eq:6}) then directly leads to
$$
{\pi M \gamma \over 2} \sum_{j=0,L/2}{\,\rm \,str\,}
(T_j + T_{j}^{-1})
$$
for the contribution of the coupling term. Combining terms and
taking the continuum limit, we finally obtain the two expressions
$S_z$ and $S_{\rm T}$ of Eq. (\ref{eq:4}) for the contribution of
energy and coupling operators to the effective action, respectively.

\subsection{Integration over massive modes}
\label{sec:integr-over-mass}

We finally turn to the discussion of the role played by the massive
modes $U$. First, notice that due to the presence of a weight term
$\sim \exp(-{N\over 2}\sum {\,\rm str\,}(Q^2 + P^2))$ in the action and
$$
N{\rm str\,}(Q^2 + P^2) = N{\rm str\,}((Q+iP)(Q-iP)) = -{N\lambda^2\over
  8}{\rm\,str\,}(U^2)
$$
fluctuations of these fields are strongly inhibited. Starting out
from an ansatz $U = \exp i W$, where $W$ is some generator, we may thus
perturbatively expand the action around $W=0$. The actual
realization of this program is cumbersome (cf.
\cite{altland99:NPB_flux} for a concrete example.) Since the result,
an extra Goldstone mode operator weakly coupled to the action, will
not play much of a role in the present analysis we restrict ourselves to
a schematic outline of the calculation.

Perturbative expansion of the 
action in powers of $W$  obtains an expression like
\begin{eqnarray*}
&&Z[0] = \int {\cal D}T e^{-S[T]} 
\left\langle e^{-S^{(1)}[T,W]+S^{(2)}[T,W]+\dots}\right\rangle_W\approx\\  
&&\qquad \approx
\int {\cal D}T e^{-S[T]} 
 e^{-\langle S^{(2)}[T,W]\rangle_W + {1\over 2}\langle
   (S^{(1)}[T,W])^2\rangle_W } 
\end{eqnarray*}
where $\langle \dots \rangle_W \equiv \int {\cal D}W
e^{-{N\lambda^2\over 8}\str(W^2)}$ and $S^{(n)}[T,W]$ denotes the
expansion of the action to $n$-th order in $W$. The ellipses stand for
contributions of higher order in $W$ which can safely be neglected
(due to the large overall factor $N\gg 1$). It is straightforward to
verify that the explicit evaluation of the operators $S^{(n)}[W,T]$
obtains contributions of the structure $ c_1 N\str(W \Phi)$ and $c_2
N\str(W \Phi W \Phi)$, where $\Phi \equiv T \partial T^{-1}$ and the
coupling constants $c_{1,2}$ are proportional to powers of $\lambda$.
Substituting these expressions back into the action and performing the
Gaussian integration over $W$, we arrive at
\begin{eqnarray*}
&&  \langle \str(S^{(1)}[T,W])^2\rangle_W \propto  N c_1^2
  \str(\Phi^2) = N c_1^2 \str(\partial T \partial T^{-1}),\\
&&  \langle \str( S^{(2)}[T,W])\rangle_W \propto  c_2
  \left[\str(\Phi)\right]^2 =  c_2 
\left[\str(T \partial T^{-1})\right]^2.
\end{eqnarray*}
What can be said about structure and relevance of these operators?
First, the two expressions $\str(\Phi^2)$ and $\str^2(\Phi)$ are the
only invariant Goldstone mode operators with two derivatives.  The
first of these is already contained in the action (cf.  Eq.
(\ref{eq:4})) with a coupling constant parametrically larger than the
constant $N c_1^2$ obtained above. Thus, the first of the two
contributions coming from the massive mode integration is irrelevant.
In contrast, the second contribution must be taken seriously and,
after integration over spatial coordinates, gives the term $S_{\rm
  Gade}$ of Eq. (\ref{eq:4}).

This completes our derivation of the effective action of the model.
Finally, to compute the correlation functions $C^{(1,2)}$ defined in
the text, we have to add the source field $\hat J$ to the partition
function, expand to first or second order and differentiate.  The
structure of the resulting expressions depends on the index
configuration of the correlation functions. For the correlators
relevant to the computation of conductance and DoS, respectively, we
obtain
\begin{eqnarray}
&&  C^{(1)}_{\alpha\alpha'} = -{i\over 2}\delta_{\mu\mu'} \left\{
    \begin{array}{ll}
\left\langle T_{{l\over2},11}\right \rangle &,\qquad l\;\mbox{even}\\
\left\langle T^{-1}_{{l-1 \over 2},11} \right\rangle 
&,\qquad l\;\mbox{odd} 
    \end{array}\right.,\nonumber\\
&&  C^{(2)}_{\alpha\alpha'\alpha'\alpha} = -{1\over 4}\delta_{\mu\mu'} \left\{
    \begin{array}{ll}
\left\langle T_{{l\over2},12} T_{{l'\over2},21} \right \rangle  &,\qquad
l\;\mbox{even},\;l'\;\mbox{even}\\ 
- \left\langle T_{{l\over2},12}T^{-1}_{{(l'-1)\over2},21}\right
\rangle  &,\qquad
l\;\mbox{even},\;l'\;\mbox{odd}\\ 
- \left\langle T^{-1}_{{(l-1)\over 2},12}T_{{l'\over2},21} \right
\rangle  &,\qquad
l\;\mbox{odd},\;l'\;\mbox{even}\\ 
\left\langle T^{-1}_{{(l-1)\over2},12}T_{{(l'-1)\over2},21}\right
\rangle  &,\qquad
l\;\mbox{odd},\;l'\;\mbox{odd} 
    \end{array}\right.,
\label{eq:10}
\end{eqnarray}
where the angular brackets stand for the functional average.

\section{Geometry of $\GL$}
\label{sec:geometry-gl}

The canonical metric on the supergroup $\GL$ derives from the
differential two-form $\omega \equiv - \str(dT dT^{-1})$. $\omega$ is
not a {\it positive} two-form, but its restriction to the sub-manifold
${\cal M}\subset \GL$ defined in the text is; ${\cal M}$ is the
maximally {\it Riemannian} subset of $\GL$.  Parameterizing the group
manifold in terms of some coordinates, $T=T(x_1,\dots,x_4)$ (half of
which are anti-commuting), the metric two-form assumes the form
$$
\omega = \sum_{ij} g_{ij} dx_i dx_j,
$$
where $\{g_{ij}\}$ defines the metric tensor (represented in the
basis $\{ x_i\}$.) As with non-super Riemannian manifolds, the metric
tensor determines the geometry of the manifold. Specifically, the
invariant group integral is defined through
$$
\int dT = \int \prod_i dx_1 \;J(x_1,\dots,x_4),
$$
where the Jacobian $J\propto  {\rm sdet} g$. Similarly, the Laplacian has
the standard structure
$$ 
\Delta = {1\over g^{1/2}}\partial_{x_i} g^{ij} g^{1/2} \partial_{x_j},
$$
where $g^{ij}$ is the inverse of $g$, $g_{ij}g^{jk}=\delta_{i}^k$. 

We next wish to represent the metric tensor, the invariant measure,
and the Laplacian in the polar decomposition, $T=kak^{-1}$ defined in
Eq.  (\ref{eq:8}).  Due to the manifest invariance of $\omega$ under
transformation $T\to k_0 T k_0^{-1}$, $k_0$ a fixed rotation matrix,
it is sufficient to evaluate the metric tensor at $k=\openone$.
Substitution of the decomposition $T=k a k^{-1}$ into the two form
then yields the covariant structure
\begin{eqnarray*}
  &&\omega = - \str\left.\left(d(k a k^{-1}) d(ka^{-1} k^{-1})\right)\right|_{k=\openone} = 
- \str\left(([dk,a] + da)\;([dk,a^{-1}]+da^{-1})\right)=\\
&&\hspace{2.0cm}- \str\left([dk,a][dk,a^{-1}] + dada^{-1}\right).
\end{eqnarray*}
Using that
$$
dk = \left(\matrix{&d\eta\cr d\nu &}\right),\qquad 
da = \left(\matrix{e^x dx&\cr & ie^{iy}dy}\right)
$$
it is straightforward to verify by elementary matrix manipulation
that the metric bilinear form
assumes the form
$$
{\bf g} = \left(\matrix{1&0&&\cr
0&1&&\cr
&&0& -4 \sinh^2({x-iy\over 2})\cr
&& +4 \sinh^2({x-iy\over 2})&0}\right),
$$
where ${\bf g} = \{g_{ij}\}$, a vectorial structure $d{\bf
  x}=(dx,dy,d\eta,d\nu)$ is understood and empty blocks are filled
with zeros. The associated superdeterminant,
$$
g= {1\over \left(4\sinh^2\left({x-iy\over 2}\right)\right)^2}.
$$
From $g$ derives the unit-normalized group integral
\begin{equation}
  \label{eq:22}
  \int dT f(T) = \int_{-\infty}^{\infty} dx\int_0^{2\pi} dy \int d\eta
  d\nu  J(x,y) f(x,y,\eta,\nu),
\end{equation}
with Jacobian
\begin{equation}
  \label{eq:28}
  J(x,y) = \sinh^{-2}\left({x-iy\over 2}\right).
\end{equation}
Further, the {\it radial} part of the Laplacian reads as
$$
\Delta = \sum_{i=x,y} g^{-1/2} \partial_i g^{1/2} \partial_i.
$$

\section{Conductance of Short Systems}
\label{sec:cond-short-syst}

Consider a short sublattice system of length $L<\xi/(M\gamma)$. This
is the quantum dot regime, where the conductance is not Ohmic but
rather determined by the coupling of the system to the leads. For a
normal, non-sublattice system, $g\sim M \gamma$, reflecting that each
of the $M$ channels contributes to transport with an efficiency set by
the coupling. We here wish to explore to which extent this behaviour
generalizes to the sublattice system.  Specifically, three different
cases will be discussed: (a) smoothened coupling where a set of
sites at each end is connected -- the type of coupling considered in
the text, (b) coupling through a single site on either end where the
two terminal sites are of the same parity, odd/odd, say.  (c) One of
the two terminating sites is even, the other odd,

That the system is short means that the functional integral is
controlled by spatially uniform configurations $T={\rm const.}$; the
stiffness introduced by the gradient term is too strong to allow for
significant fluctuations. (More precisely, fluctuating field
configurations lead to relative corrections of ${\cal
  O}(LM\gamma/\xi)$  which we are not going to consider.)

Let us begin by considering case (a).  Evaluation of the functional
expectation value (\ref{eq:12}) for a spatially uniform field
configuration leads to the expression
\begin{eqnarray*}
  g^{(a)}= -\left({M\pi\gamma\over 2}\right)^2 \int dT 
  (T-T^{-1})_{12}(T-T^{-1})_{21} e^{-\pi M \gamma \str(T+T^{-1})}. 
\end{eqnarray*}
We next need to do the group integral. Inspection of the exponent
shows that the integration is dominated by configurations close
to the group origin, $T=\openone$. This suggests to represent our
$T$'s as $T=\exp W$ and to integrate over the generators $W$ in a
Gaussian approximation:
\begin{eqnarray*}
  g^{(a)}\approx -(M \pi\gamma)^2 \int dW
  W_{12}W_{21} e^{-\pi M \gamma\str W^2},
\end{eqnarray*}
where we have used that close to the group origin the integration measure
is flat. Doing the integral over the components of $W$ one
finds
$$
g^{(a)}= {M\pi \gamma \over 2} 
$$
in agreement with the behaviour of a non-sublattice
system. The situation in case (b) is not much different. Noticing that
fields $T$ sit at the odd sites of our system, while $T^{-1}$'s are
attached to even sites, we find that the conductance is expressed
as
\begin{eqnarray*}
  g^{(b)}= (M\pi\gamma)^2 \int dT \;
  T_{12}(T^{-1})_{21} e^{-\pi M \gamma \str(T+T^{-1})}. 
\end{eqnarray*}
(The disappearance of the combination $T-T^{-1}$ in the pre-exponent
reflects the fact that only {\it single} sites of different parity are
coupled to the continuum. Again the integral is dominated by the group
origin and similar reasoning as above leads to
\begin{eqnarray*}
  g^{(b)}\approx -\left({M \pi\gamma\over 2}\right)^2 \int dW
  W_{12}W_{21} e^{-{\pi M \gamma \over 4}\str W^2}= {M\pi \gamma\over 2}.
\end{eqnarray*}
In case (c), however, the situation is different. Owing to
the identical parity of the terminating sites, the integral
representation of the conductance now assumes the form 
\begin{eqnarray*}
  g^{(c)}= -(M\pi\gamma)^2 \int dT \;
  T_{12}T_{21} e^{-\pi M \gamma \str(T^2)}. 
\end{eqnarray*}
This is an unpleasant expression: the exponential weight no longer
projects onto the group origin, in fact does not even have a stable
saddle point. To compute the conductance we therefore have to
integrate over the full group manifold, a task that is most
efficiently done in polar coordinates. Using Eqs. (\ref{eq:8}) and
(\ref{eq:22}) and integrating out Grassmann components it is
straightforward to verify that the radial part of the integral assumes
the form
$$
g^{(c)} =(M\pi\gamma)^2 \int_{-\infty}^{\infty} dx
\int_0^{2\pi} \; {(e^{x}-e^{iy})^2\over \sinh^2\left({x-iy\over
      2}\right)}
e^{-(e^x-e^{iy})}=0. 
$$
To understand the vanishing of the integral, notice that $\int dy =
{1\over 2\pi i} \oint {dz\over z}$ can be transformed into the
integral of the complex variable $z=e^{iy}$ over the unit circle. One
verifies that, regardless of the value of $x$, the integrand is
analytic and void of singularities inside the integration contour.
Cauchies theorem then implies vanishing of the integral.


\begin{references}
\bibitem{fn2}
Note that the simplictiy inherent to the radial sector of the heat equation is
  not specific to the manifold $\GL$. In fact, it is a generic feature of {\it
  all} group valued supermanifolds that the radial components of their
  Laplacians can be reduced to the form $\Delta + C$, where $\Delta$ is flat
  and $C$ some constant\protect\cite{berezin}. In our case $C=0$ which {\it is} special for ${\rm
  GL}(n|n)$.).
\end{references}

\end{document}